\documentstyle[aps,epsf,rotate]{revtex}
\font\mb=msbm10
\font\helv=cmss10
\def\bPhi{{\bf\Phi}}

\begin{document}
\bibliographystyle{unsrt}
\draft

\title{Chaotic and fractal properties of deterministic diffusion-reaction
processes} 
\author{P. Gaspard and R. Klages}
\address{Center for Nonlinear Phenomena and Complex Systems\\ 
and Service de Chimie Physique,\\
Facult\'e des Sciences, Universit\'e Libre de Bruxelles,\\
Campus Plaine, Code Postal 231, B-1050 Brussels, Belgium}
\date{\today}
\maketitle
\begin{abstract}

We study the consequences of deterministic chaos for
diffusion-controlled reaction. As an example, we analyze a
diffusive-reactive deterministic multibaker and a parameter-dependent
variation of it. We construct the diffusive and the reactive
modes of the models as eigenstates of the Frobenius-Perron
operator. The associated eigenvalues provide the dispersion relations
of diffusion and reaction and, hence, they determine the reaction
rate. For the simplest model we show explicitly that the reaction rate
behaves as phenomenologically expected for one-dimensional
diffusion-controlled reaction. Under parametric variation, we find
that both the diffusion coefficient and the reaction rate have
fractal-like dependences on the system parameter.

\end{abstract}


{\bf Matter is most often the stage of reactions which evolve on a
reactive time scale being intermediate between the long time scale of
hydrodynamic transport phenomena and the short time scale of
microscopic chaos.  This chaos is generated by the collisions between
the atoms and molecules of the fluid beyond a temporal horizon caused
by the Lyapunov instability of motion. Under nonequilibrium
conditions, long-time trajectories organize themselves in phase space
to form fractal structures and weird invariant or conditionally
invariant measures, the consequences of which have just started to be
explored.  In this perspective, we study here simple models of
diffusion-reaction processes in order to confront the phenomenology
with the new approach based on deterministic chaos.}

\section{Introduction}
Irreversible phenomenological equations such as the Navier-Stokes or
diffusion-reaction equations describe transport and reaction processes
in fluid flows or chemical reactions occurring on macroscopic
spatio-temporal scales. For instance, spatial inhomogeneities of size
$\cal L$ in the density are damped by diffusion over a time scale of
the order of $T_{\rm diff}\sim {\cal L}^2/D$, where $D$ is the
diffusion coefficient.  In typical laboratory experiments, the size
$\cal L$ of the inhomogeneities is of the order of millimeters,
centimeters, or larger so that the relaxation time $T_{\rm diff}$ is
macroscopic.

Recent works have studied the relationship between these macroscopic
processes and the much faster process of chaos in the microscopic
motion of atoms or molecules in the fluid
\cite{Dorfman,Hoover,PGbook,Evans,GN90,RKD,BTV}.
The defocusing character of the collisions between the particles of
the fluid is at the origin of a very high dynamical instability in the
microscopic motion. This instability, which is a major phenomenon at
the center of the current interest, is characterized by a spectrum of
positive Lyapunov exponents, as shown by many recent numerical and
analytical studies \cite{Livi,Posch,vanBeijerenD}. The maximum
Lyapunov exponent of a dilute gas of particles of diameter $d$ is
typically of the order of
\begin{equation}
\lambda_{\rm max} \ \sim \ \frac{v}{\ell} \ \ln \frac{\ell}{d} \ , 
\end{equation}
where $\ell$ is the mean free path between the collisions, and $v$ is
the mean velocity of the particles \cite{Krylov,GSoli}. Consequently,
the time scale over which the dynamical instability develops in the
motion of a particle due to the collisions with the surrounding
particles is of the order of the inverse of the maximum Lyapunov
exponent which takes the value $T_{\rm chaos} \sim \lambda_{\rm
max}^{-1} \sim 10^{-10}$ sec for a gas at room temperature and
pressure.

The microscopic time scale is in contrast with the macroscopic time
scale of transport and reaction processes. This presence of two
different time scales is an essential feature in the recent
establishment of quantitative relationships between macroscopic
transport and microscopic chaos in both the thermostatted-system and
the escape-rate approaches
\cite{Hoover,PGbook,Evans,GN90,Posch,GB95,DG95,GD95}. Indeed, in both
approaches transport coefficients are related to
{\em differences} between two characteristic quantities of chaos: The
thermostatted-system method works with the difference between the
maximum and the absolute value of the minimum Lyapunov exponent
\cite{Hoover,Evans,Posch}, whereas the escape-rate approach employs the
difference between the positive Lyapunov exponents and the
Kolmogorov-Sinai entropy \cite{PGbook,GN90,GB95,DG95,GD95}. In this
sense, the nonequilibrium property of transport is related to a slight
disbalance between the dynamical instability, which is the cause, and
the induced temporal disorder, which is the effect. At equilibrium,
the effect exactly compensates the cause. Away from equilibrium,
temporal disorder is slightly reduced to the benefit of transport,
which appears as an extra effect of the dynamical instability beside
the temporal disorder. In this scheme, the fluid appears at the stage
of a highly chaotic motion of its constituent particles, which
animates different possible transport processes if the system is
maintained out of equilibrium.

The hypothesis of microscopic chaos \cite{GSoli} or the chaotic hypothesis
\cite{Gallav} replace the old stochastic hypotheses in nonequilibrium
statistical mechanics. Previously, the stochastic hypotheses assumed
that transport processes arise from stochastic effects, such as the
Langevin white noise which has an infinite Kolmogorov-Sinai entropy
per unit time. However, these processes have thus a much larger
temporal disorder than allowed by Newton's deterministic equations of
motion \cite{GAdv97}. The new chaotic hypothesis has the enormous
advantage of assuming a temporal disorder which is now compatible with
the determinism of a microscopic Newtonian dynamics.  In this regard,
the chaotic hypothesis overwhelms the previous stochastic hypotheses,
which nevertheless remains of great usefulness in their domain of
validity.

The purpose of the present paper is to describe several consequences
of the hypothesis of microscopic chaos in diffusion-reaction
systems. This class of physico-chemical processes has not yet been
explored in the perspective of understanding their kinetics on the
sole assumption of microscopic chaos, without involving stochastic
Langevin or birth-and-death processes. The diffusion-reaction systems
are particularly important in various fields of chemical physics, such
as chemical kinetics, homogeneous and heterogeneous catalyses, pattern
formation in nonequilibrium reactions and in morphogenesis,
recombination processes in solid or liquid phases, as well as
high-energy reaction processes in astrophysical systems \cite{GP71}.

We shall focus here on the simplest diffusion-reaction process with
a linear chemical reaction law,
\begin{equation}
A \ \rightleftharpoons \ B \ , \label{reaction}
\end{equation}
which already provides a nontrivial dynamics. As a vehicle of our
study we shall use multibaker models, which are deterministic
versions of discrete Markov processes. The deterministic dynamics of a
multibaker has therefore the same finite and positive Kolmogorov-Sinai
entropy per unit time as the corresponding discrete Markov
process. The models we propose are spatially extended generalizations
of a baker-type model of isomerization previously studied by Elskens
and Kapral \cite{Elskens}.

This paper is organized in the following way: In Section II, we
construct multibaker models of diffusion-reaction by starting from a
diffusive-reactive Lorentz gas. In Section III, we focus on the
diffusive properties of the multibakers. By employing quasiperiodic
boundary conditions we show that for the simplest model the diffusive
properties are the same as those of the previously discussed dyadic
multibaker of diffusion. We then demonstrate that under parametric
variation of this model the diffusion coefficient exhibits a
self-similar structure reminiscent of fractal curves. In Section IV,
we describe the reactive properties. With quasiperiodic boundary
conditions, we first study the simplest model and derive the
dispersion relation of the chemiodynamic modes. We explicitly
construct the phase-space distribution of these modes and define the
reaction rate by comparison with phenomenology. We then show that the
reaction rate behaves in a highly irregular manner if we consider the
parameter-dependent model. Conclusions are drawn in Section V.

\section{Deterministic models of diffusion-reaction processes}
\subsection{Definition of the models}
\begin{figure}
\epsfxsize=18cm
\centerline{\epsfbox{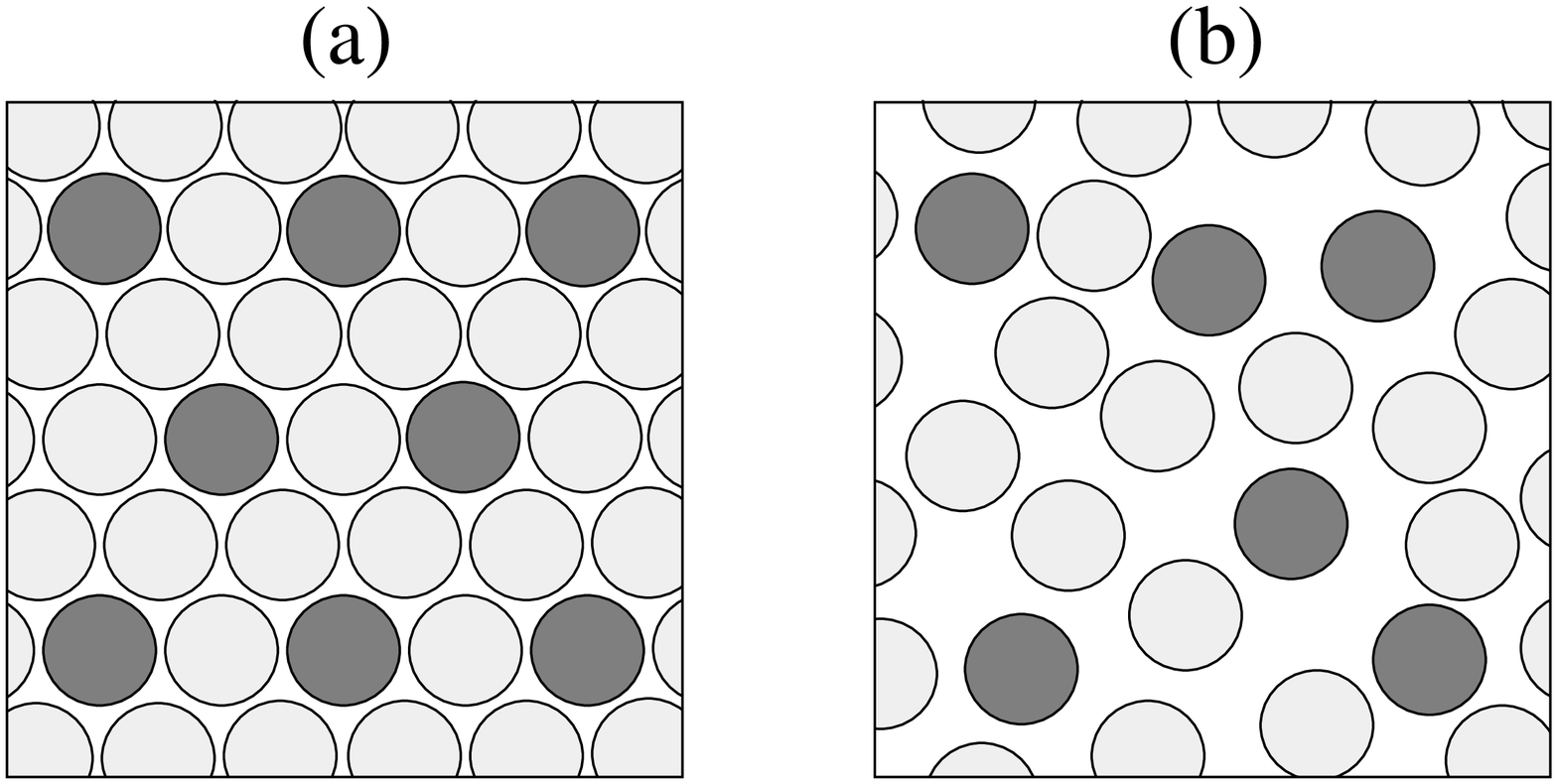}}
\caption{\label{figLorentz} Examples of reactive Lorentz-type models:
(a) on a regular lattice; (b) on a random lattice.}
\end{figure}

In order to motivate the introduction of the multibaker models, we
first consider a reactive Lorentz gas in which a point particle
undergoes elastic collisions on hard disks which are fixed in the
plane. The disks may form a random or a regular configuration (Fig.\
\ref{figLorentz}).  A fraction of the disks are supposed to be
catalysts where the point particle changes its state, or color, from
$A$ to $B$ or vice versa at the instant of the collision.  The mass of
the particle is assumed to be the same in both states $A$ and $B$.
The phase space coordinates of each particle are given by its
position, its velocity, and its color $(x,y,v_x,v_y,c)$ with
$c\in\lbrace A,B\rbrace$. Since energy is conserved at the elastic
collisions, the magnitude of the velocity is a constant of motion,
$v=\sqrt{v_x^2+v_y^2}$, so that the coordinates reduce to
$(x,y,\varphi,c)=({\bf X},c)$, where $\varphi={\rm arctan}(v_y/v_x)$
is the angle between the velocity and the $x$-axis.

The motion induces a time evolution of the phase-space probability
densities, or concentrations, for each color,
\begin{equation}
{\bf f}({\bf X}) \ = \ \pmatrix{ f({\bf X},A)\cr f({\bf X},B)\cr} \ .
\label{density} \end{equation}
The mean phase-space density, defined by the average of the
concentrations, does not distinguish between the colors. Thus, we may
expect that
\begin{equation}
\tilde f({\bf X}) \ = \ \frac{1}{2}\lbrack f({\bf X},A) + f({\bf X},B)\rbrack
 \label{mean}  \end{equation}
evolves in time exactly as in the non-reactive Lorentz gas, as studied
elsewhere \cite{BunimSinai,Chernov,PRE96}. The dynamics of reaction
should appear in the difference between the concentrations
\begin{equation}
g({\bf X}) \ = \  f({\bf X},A) - f({\bf X},B) \label{difference}
\end{equation}
which is expected to follow a macroscopic relaxation toward zero if
there is equipartition of particles between both colors.

It has been explained elsewhere that the flow dynamics of the Lorentz
gas can be reduced to a Birkhoff mapping from collision to collision
\cite{PRE96}. Each collision can be represented by two variables: the
angle $\theta$ giving the position of impact on the perimeter of the
disk as $(x=\cos\theta,y=\sin\theta)$ and the angle $\phi$ between the
velocity after the collision and the normal at impact.  The sine of
the velocity angle $\varpi=\sin\phi$ together with the position angle
$\theta$ are the so-called Birkhoff coordinates, in which the mapping
is area-preserving. All the collision events with the disk of label
$l$ are thus represented by the rectangle
\begin{equation} 
{\cal R}_l \ = \ \bigl\lbrace
(\theta,\varpi,l): \: 0 \leq \theta <2\pi\ , \: -1\leq \varpi \leq +1
\bigr\rbrace \ . \label{rect} 
\end{equation}
The dynamics of collisions can therefore be written as the Birkhoff mapping
\begin{equation}
(\theta_{n+1},\varpi_{n+1},l_{n+1},c_{n+1}) \ = \
\bPhi(\theta_{n},\varpi_{n},l_{n},c_{n}) \ , \label{Birkh}
\end{equation}
which is known to be area-preserving, time-reversal symmetric, and of
hyperbolic character.

\begin{figure}
\epsfxsize=14cm
\centerline{\epsfbox{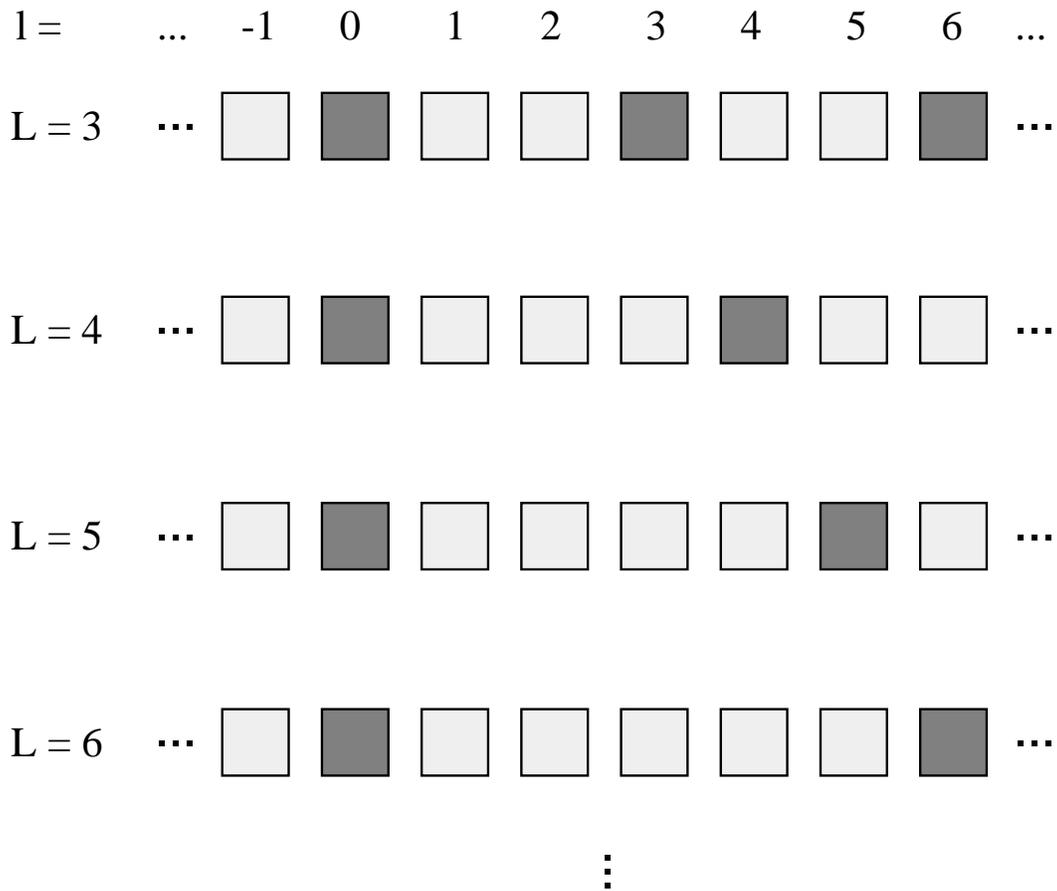}}
 
\vspace*{0.3cm}
\caption{\label{figBakery} Geometry of a reactive multibaker or bakery
map.}
\end{figure}

A caricature of this mapping is provided by a multibaker model (Fig.\
\ref{figBakery}) in which we suppose that the rectangular domains Eq.\
(\ref{rect}) representing the disks are replaced by squares
\begin{equation}
{\cal S}_l \ = \ \bigl\lbrace (x,y,l): \: 0 \leq x \leq 1\ ,
\: 0\leq y \leq 1 \bigr\rbrace \ , \label{square}
\end{equation}
where $l\in\hbox{\mb Z}$ is the label of the square.  Each square of
the multibaker model corresponds to a disk of the Lorentz gas.  Now,
the collision dynamics is simplified by replacing the complicated
Birkhoff map Eq.\ (\ref{Birkh}) by a baker-type map with horizontal
stretching by a factor of two, followed by cutting the elongated
square into two.  The collisions from disk to disk are replaced by
jumps of the particle from square to square according to the
transition rule $l\to l-1$ if $x\leq 1/2$ and $l\to l+1$ if $x>1/2$
between next-neighboring squares.  The squares are arranged such that
they form a one-dimensional chain.  One out of $L$ squares is assumed
to be a catalyst where the color changes from $c=A$ (resp. $B$) to its
complement $\bar{c}=B$ (resp. $A$). The map of the model is thus
\begin{equation}
\bPhi(x,y,l,c) \ = \ \cases{ \Bigl(2x,\frac{y}{2},l-1,c'\Bigr) \ ,
\quad 0\leq x
\leq \frac{1}{2} \ , \cr \cr \Bigl(2x-1,\frac{y+1}{2},l+1,c'\Bigr) \ , \quad
\frac{1}{2}< x \leq 1 \ , \cr}\label{multibaker}
\end{equation}
where $c'=\bar{c}$ if $l=0,\pm L,\pm 2L,...$ and $c'=c$ otherwise.
This map is area-preserving, time-reversal symmetric, and chaotic with
a positive Lyapunov exponent $\lambda_+=\ln 2$ and a negative one
$\lambda_-=-\ln 2$, such as the multibaker map
\cite{PG1,Gasp93,Gasp94,TG1}.

In our model, the reaction is controlled by the diffusion if the
reactive sites are diluted in the system.  This important case of
chemical reactions has been much studied in the literature since
Smoluchowski seminal work \cite{Smoluchowski}. It is known that the
macroscopic reaction rate is determined by the time taken by particles
to diffuse toward the reactive site. A crossover occurs at dimension
two which is the Hausdorff dimension of a Brownian path. Accordingly,
the flux of reactants toward a catalyst is sensitive to the presence
of the next-neighboring catalysts in less than two dimensions, but not
in systems of dimensions higher than two. In particular, in a
one-dimensional system the reaction rate should behave as
\begin{equation}
\kappa \ \sim \ \frac{D}{L^2} \ , 
\label{DiffControlReact} \end{equation}
where $D$ is the diffusion coefficient and $L$ is the distance between
the reactive sites or catalysts.

A main goal of our work is to investigate the dynamical properties of
our diffusion-reaction model in order to know whether this expected
macroscopic behavior is confirmed from the microscopic dynamics or
not. We shall also consider a parametric variation of this model with
a more complicated dynamics. This is due to an extra dependency on a
shift parameter which is introduced when the half squares are glued
back into the chain. When this continuous parameter varies it induces
topological changes in the trajectory dynamics which are reminiscent
of the topological changes induced by varying the disk radius in the
Lorentz gas \cite{BunimSinai,Chernov,PRE96}. The parametric extension
of the multibaker model shows that the diffusion coefficient as well
as the reaction rate may vary in a highly irregular fashion as a
function of a parameter. This is an important consequence of
deterministic chaos which already appears on the level of
one-dimensional maps, as will be discussed in the following section.

\subsection{The Frobenius-Perron operator and quasiperiodic boundary
conditions}
Since Boltzmann's work, it is well known that transport and
reaction-rate processes should be conceived in a statistical sense
because the individual trajectories are affected by the famous
Poincar\'e recurrences.  We therefore consider the time evolution of
statistical ensembles of trajectories as represented by the
probability densities Eq.\ (\ref{density}). They evolve in time
according to the Frobenius-Perron equation
\begin{equation}
f_{t+1}(x,y,l,c) \ = \ f_t\Bigl\lbrack \bPhi^{-1}(x,y,l,c)\Bigr\rbrack
\ \equiv \ (\hat P f_t)(x,y,l,c) \qquad (t\in\hbox{\mb Z}) \ .
\end{equation}
We choose quasiperiodic boundary conditions by assuming that the
solution of the Frobenius-Perron equation is quasiperiodic on the
chain with a wavenumber $k$.  Moreover, we suppose that the solution
decays exponentially with a decaying factor $\chi=\exp s$ where
$\vert\chi\vert \leq 1$ or ${\rm Re}\ s\leq 0$,
\begin{equation}
f_t(x,y,l,c) \ \sim \ \chi^t \ \exp(ikl) \ .
\end{equation}
The decay rate $s$ is calculated by solving the eigenvalue problem of
the Frobenius-Perron operator. We note that the Frobenius-Perron
operator is in general non-unitary so that root vectors associated
with possible Jordan-block structures may exist beyond the
eigenvectors.  We shall focus here on the eigenvectors because they
control the slowest decay on the longest time scales
\cite{Gasp93,Gasp94}.

For quasiperiodic solutions, the Frobenius-Perron operator reduces to
the following Frobenius-Perron operator $\hat Q_k$ which depends on
the wavenumber $k$ and acts on functions which are defined only in $L$
successive squares of the chain,
\begin{equation}
\hat Q_k \ \equiv \cases{
f_{t+1}(x,y,0,c) \ = \ \theta\Bigl(\frac{1}{2}-y\Bigr) \
f_t\bigl(\frac{x}{2},2y,1,c\bigr) \ + \ e^{-ikL} \
\theta\Bigl(y-\frac{1}{2}\Bigr) \ f_t\bigl(\frac{x+1}{2},2y-1,L-1,c\bigr)\ , 
\cr
f_{t+1}(x,y,1,c) \ = \ \theta\Bigl(\frac{1}{2}-y\Bigr) \
f_t\bigl(\frac{x}{2},2y,2,c\bigr) \ + \ \theta\Bigl(y-\frac{1}{2}\Bigr) \
f_t\bigl(\frac{x+1}{2},2y-1,0,\bar{c}\bigr)\ , \cr
f_{t+1}(x,y,2,c) \ = \ \theta\Bigl(\frac{1}{2}-y\Bigr) \
f_t\bigl(\frac{x}{2},2y,3,c\bigr) \ + \ \theta\Bigl(y-\frac{1}{2}\Bigr) \
f_t\bigl(\frac{x+1}{2},2y-1,1,c\bigr)\ , \cr
\qquad\qquad\vdots \cr
f_{t+1}(x,y,L-2,c) \ = \ \theta\Bigl(\frac{1}{2}-y\Bigr) \
f_t\bigl(\frac{x}{2},2y,L-1,c\bigr) \ + \ \theta\Bigl(y-\frac{1}{2}\Bigr) \
f_t\bigl(\frac{x+1}{2},2y-1,L-3,c\bigr)\ , \cr
f_{t+1}(x,y,L-1,c) \ = \ e^{ikL} \ \theta\Bigl(\frac{1}{2}-y\Bigr) \
f_t\bigl(\frac{x}{2},2y,0,\bar{c}\bigr) \ + \ 
\theta\Bigl(y-\frac{1}{2}\Bigr) \
f_t\bigl(\frac{x+1}{2},2y-1,L-2,c\bigr)\ . \cr}
\end{equation}
We notice that there is a reaction, i.e., a change of color, for
particles passing the cell $l=0$ so that a concentration with $l=0$
and $\bar{c}$ appears in the second and in the last line.  We also
notice that there is a factor $\exp(-ikL)$ in the first line for the
particle coming from the previous segment of length $L$ in the
infinite chain, where the concentration functions are multiplied by
the factor $\exp(-ikL)$.  On the other hand, there is a factor
$\exp(ikL)$ in the last line for the particle coming from the next
segment where the concentration functions are multiplied by
$\exp(ikL)$.  Otherwise, this Frobenius-Perron operator is the same as
in the infinite dyadic multibaker model studied in Refs.\
\cite{PG1,Gasp93,Gasp94}.

As we discussed in the previous section, the presence of two chemical
components $c=A$ or $B$ implies that the Frobenius-Perron operator
$\hat Q_k$ acts on $2L$ functions which can be linearly combined to
separate the functional space in two subspaces on which two decoupled
Frobenius-Perron operators would act.  The first subspace is defined
by Eq.\ (\ref{mean}) where the Frobenius-Perron operator reduces to
the diffusive Frobenius-Perron operator of the multibaker map. The
second subspace is defined by Eq.\ (\ref{difference}) which gives a
different evolution operator of reactive type. Diffusive properties
are studied in the next Section III, while reactive properties will be
discussed in Section IV.

\section{Diffusion dynamics}
\subsection{Diffusive modes of the dyadic multibaker}
In this subsection, we consider the diffusive dynamics of the dyadic
multibaker model Eq.\ (\ref{multibaker}) with quasiperiodic boundary
conditions.  The subspace of diffusion is defined by the mean density
of Eq.\ (\ref{mean}). For hydrodynamic modes of wavenumber $k$ we thus
write
\begin{equation}
\tilde f_t(x,y,l) \ = \ \frac{1}{2}\bigl\lbrack
f_t(x,y,l,A)+f_t(x,y,l,B)\bigr\rbrack \ \equiv \ \exp\Biggl\lbrack i\biggl(
k+2\pi \frac{\nu}{L}\biggr) l  \Biggr\rbrack \ \eta_t(x,y) \ ,
\end{equation}
and the new function obeys the simpler evolution equation
\begin{equation}
\eta_{t+1}(x,y) \ = \ (\hat Q_k^{({\rm D})} \eta_t)(x,y) \ \equiv \
e^{+i(k+2\pi\nu/L)} \ \theta\biggl(\frac{1}{2}-y\biggr) \
\eta_t\biggl(\frac{x}{2},2y\biggr) \ + \ e^{-i(k+2\pi\nu/L)} \
\theta\biggl(y-\frac{1}{2}\biggr) \ \eta_t\biggl(\frac{x+1}{2},2y-1\biggr) \ , 
\end{equation}
which is the Frobenius-Perron equation for the dyadic multibaker map
based on quasiperiodic boundary conditions \cite{Gasp93}. The
respective Frobenius-Perron operator has been analyzed in detail
elsewhere \cite{Gasp93,Gasp94}. Its decay rates are
\begin{equation}
s_{m\nu}(k) \ = \ \ln \ \chi_{m\nu}(k) \ = \ - \ m \ \ln 2 \ + \ \ln
\cos\biggl(k+\frac{2\pi\nu}{L}\biggr) 
\end{equation}
with $m=0,1,2,3,...$, $\nu=0,1,2,...,L-1$, and with a degeneracy of
$(m+1)$. The eigenvectors
\begin{equation}
(\hat Q_k^{({\rm D})} \ \psi_{m\nu})(x,y;k) \ = \ e^{s_{m\nu}(k)} \
\psi_{m\nu}(x,y;k) \ , 
\end{equation} 
and some root vectors have been constructed in Refs.\
\cite{Gasp93,Gasp94} in terms of the cumulative functions
\begin{equation}
F_{0\nu}(x,y;k) \ = \ \int_0^x \: dx' \: \int_0^y \: dy' \
\psi_{0\nu}(x',y';k) \ . 
\end{equation}
For small enough wavenumbers $k$, these cumulative functions are
continuous functions which are products of a monomial in $x$ with a
nondifferentiable de Rham function in $y$.  Accordingly, the
eigenvectors $\psi_{0\nu}$ are complex singular measures for small
enough $k$.

We observe that the decay rate with $m=0$ and $\nu=0$ vanishes
quadratically as $k\to 0$ in agreement with the expected diffusive
behavior,
\begin{equation}
s_{00}(k) \ = \ \ln\ \cos \ k \ = \ - \ \frac{k^2}{2} \ - \ \frac{k^4}{12} \
\cdots \ , \label{disp}
\end{equation}
which shows that the diffusion coefficient is $D=1/2$.

The nonequilibrium steady states corresponding to a uniform density
gradient $g$ for the mean density of Eq.\ (\ref{mean}) have also been
constructed \cite{TG1}.  It has been shown that the nonequilibrium
steady state of the infinite system corresponds to a singular
invariant measure represented by a continuous but nondifferentiable
cumulative function
\begin{equation}
F_{\rm st. \ st.}(x,y,l) \ = \ \int_0^x \: dx' \: \int_0^y \: dy' \:
\tilde f_{\rm st. \ st.}(x',y',l) \ = \ g \ l \ x \ y \ + \ g \ x \ T(y) \ ,
\label{CumulStSt}
\end{equation}
where $T(y)$ is the Takagi function obtained as a solution of the
iteration
\cite{Hata} 
\begin{equation}
T(y) \ = \ \cases{\frac{1}{2}\ T(2y) \ + \ y \ , \qquad 0 \leq y \leq
\frac{1}{2} \ , \cr\cr \frac{1}{2}\ T(2y-1) \ - \ y \ + \ 1 \ , \qquad
\frac{1}{2} < y \leq 1 \ . \cr} \label{Takagi}
\end{equation}
The Takagi function is nondifferentiable because its formal derivative
is infinite almost everywhere. It is given by a Lebowitz-McLennan type
of formula \cite{Lebowitz}
\begin{equation}
\frac{dT}{dy} \ = \ \sum_{t=0}^{\infty} \ j\bigl\lbrack M^t(y) \bigr\rbrack
\ , \label{DTakagi} \end{equation}
where $M(y)=(2y) $ (modulo 1), and $j(y)=\pm 1$ if $y<1/2$ or $y>1/2$,
respectively, is the {\em jump-velocity}.  The singular character of
the diffusive steady state turns out to be a general feature in
finite-dimensional deterministic chaos of large spatial extension, as
shown elsewhere \cite{PRE96,TG1}.

Moreover, this singular character of the steady state measure plays a
fundamental role in the explanation of the entropy production of
irreversible thermodynamics.  In particular, the expected entropy
production can be derived from the Takagi function in the case of the
multibaker map \cite{G97JSP}.  The presence of this singularity solves
the famous paradox of the constancy of the Gibbs entropy, which can be
set up when we do not recognize that the out-of-equilibrium invariant
measure is very different from the equilibrium one on fine scales in
phase space.  The out-of-equilibrium invariant measure becomes
singular if the nonequilibrium constraints are imposed at distances
larger than several mean free paths.  This is probably a paradoxical
aspect of the local equilibrium hypothesis that a nonequilibrium
system appears in local equilibrium on the largest scales of phase
space although intrinsic correlations exist on finer scales which are
due to the chaotic dynamics.  The singular character of the invariant
measure explains that there is an entropy production in large
nonequilibrium systems where the chaotic dynamics removes the
signature of determinism down to extremely fine scales in phase space.

\subsection{Diffusion coefficients in parameter-dependent models}
In this subsection, we describe an important consequence of
deterministic chaos which shows up when a system parameter is varied.
Under such circumstances, the diffusion coefficient of maps like a
parameter-dependent multibaker exhibits a fractal structure by
changing the parameter. In the same way as outlined in this
subsection, we will show later that the same behavior appears in the
reactive transport properties of our parameter-dependent
multibaker. We thus emphasize the striking analogies between the
diffusive and the reactive properties in regard to their parametric
sensitivity.

For this purpose, we summarize the main methods and results concerning
the parametric variation of the diffusive dynamics in multibakers and
in simple one-dimensional maps. To study parameter-dependent transport
in multibaker models like Eq.\ (\ref{multibaker}) such one-dimensional
maps are crucial, because they govern the dynamics of multibakers
projected onto the $x$-axis. We therefore start with a brief review on
parameter-dependent diffusion in one-dimensional maps. We then show
how all the methods and results obtained for one-dimensional maps
carry over to a two-dimensional parameter-dependent multibaker which
we introduce and discuss to the end of this subsection.

\subsubsection{Fractal forms in a Green-Kubo formula}
Chains of one-dimensional chaotic maps are the simplest dynamical
systems in which deterministic diffusion can be studied
\cite{SFK}. One may think of them as deterministic generalizations of
simple random walks on the real line, where the full microscopic
history of the particles is taken into account. In general, the
microscopic dynamics is affected by changes of some control
parameters. Hence, in contrast to the discussion of the previous
subsection where no parameter has been varied, we will be interested
here in the resulting parameter dependence of the diffusion
coefficient \cite{RKD,RKDiss,RKD2,RKT,RKDG}. As mentioned before, the
parameters to be considered may be physically related to, for
instance, varying the density or the shape of scatterers. An example
of such systems is the chain of piecewise linear maps depicted in
Fig.\ \ref{fig1}. This map is of the general form $x_{t+1}=M_a(x_t)$,
and it is periodic by satisfying the condition
$M_a(x+1)=M_a(x)+1$. The slope $a$ serves as the control parameter and
is trivially related to the Lyapunov exponent of the map via
$\lambda=\ln a$. The parameter-dependent diffusion coefficient can be
obtained by the Green-Kubo formula \cite{Dorfman,PG1,RKDiss,RKDG}
\begin{equation}
D(a) \equiv \left\langle j_a(x)\sum_{t=0}^\infty j_a\lbrack
M_a^t(x)\rbrack \right\rangle -\frac{1}{2}\left\langle
j^2_a(x)\right\rangle \quad , \label{eq:gk}
\end{equation}
where the average $\left\langle \cdot \right\rangle
\equiv\int_0^1dx\;\varrho_a(x)(\cdot)$ has to be taken over the invariant
probability density $\varrho_a(x)$ on the unit interval. $j_a(x)$
gives the integer number of boxes a particle has traversed after one
iteration starting at position $x$ and is thus the parameter-dependent
extension of the jump-velocity introduced in Eq.\ (\ref{DTakagi})
above. Thus, the complete microscopic dynamics is divided into two
parts in Eq.\ (\ref{eq:gk}): the {\em intra-cell dynamics}, that is,
the dynamics within a single box, which is represented by the
invariant probability density, and the {\em inter-cell dynamics} given
by the sum of the jump-velocities, which contains the history of the
particles travelling between the single boxes of the chain.

For computing the diffusion coefficient both parts can be treated
separately. The invariant probability density is obtained by solving
the Frobenius-Perron equation for the map restricted to the
unit interval, $\tilde{M}_a(x)\equiv M_a(x) \quad (\mbox{mod} \; 1)$,
\begin{equation} 
\varrho_{a,t+1}(x) = \int dz \; \varrho_{a,t}(z) \;
\delta\lbrack x-\tilde{M}_a(z)\rbrack \quad .  \label{eq:fpe}
\label{eq:fp}
\end{equation}
To do this, Eq.\ (\ref{eq:fpe}) can be written as a matrix equation,
where the Frobenius-Perron operator has been transformed into a
transition matrix \cite{RKD,PG1,RKDiss,Beck}. For maps of the type
considered here, exact transition matrices can be constructed whenever
a so-called Markov partition exists. This is the case for a dense set
of parameter values $a$ on the real line. The invariant probability
density can then be calculated either by solving the eigenvalue
problem of the transition matrix, which in simple cases can be
performed analytically, or by solving the Frobenius-Perron equation by
iterating the transition matrices numerically \cite{SFP}.

In Fig.\ \ref{fig2} (a) and (b), typical invariant probability
densities are plotted at two values of the slope. They are step
functions on the unit interval, where the regions of the functions
being piecewise constant correspond to the single cells of the
respective Markov partitions. For piecewise linear maps, the invariant
probability densities should always be step functions, although for
arbitrary parameter value they may consist of infinitely many steps
\cite{RKDiss,JG,Ers}. The sum of jump-velocities, as the second
ingredient of the Green-Kubo formula Eq.\ (\ref{eq:gk}), gives the
integer displacement of a particle after $t$ iterations starting at
initial position $x$. Since the system is chaotic, this function is
highly irregular in $x$. To deal with this quantity, it is more
convenient to define functions $T_a(x)$ via
\begin{equation}
\frac{dT_a}{dx} \equiv \sum_{t=0}^{\infty} j_a\lbrack
M_a^t(x)\rbrack \ ,\label{DTa}
\end{equation} 
which is a parametric extension of Eq.\ (\ref{DTakagi}). The functions
$T_a(x)$ now give the integral of the displacement of particles which
start in a certain subinterval, and they behave much more regular in
$x$ than the sums of jumps. Employing $T_a(x)= \lim_{t\to\infty}
T_{a,t}(x)$, it can be shown that these functions are obtained in
terms of the recursion relation
\begin{equation}
T_{a,t}(x)=t_a(x)+\frac{1}{a}\;T_{a,t-1}
\Bigl\lbrack\tilde{M}_a(x)\Bigr\rbrack
\ , \label{eq:rr}
\end{equation}
with $t_a(x)$ being determined by $dt_a/dx\equiv j_a(x)$ and by
requiring that $T_a(0)=T_a(1)=0$. $T_a(x)$ can be computed by
iterating Eq.\ (\ref{eq:rr}) numerically. For two special values of
the slope the results have been plotted in Fig.\ \ref{fig2} (c) and
(d). The functions $T_a(x)$ are self-similar on the unit interval and
scale with the slope $a$. For $a=2$, Eq.\ (\ref{eq:rr}) appears as a
special case of Eq.\ (\ref{Takagi}). Therefore, functions like
$T_a(x)$ may be denoted as {\em generalized Takagi functions}.

The numerically exact result for the parameter-dependent diffusion
coefficient is shown in Fig.\ \ref{fig3} for $2\le a\le 8$. Naively,
one may have expected that $D(a)$ increases monotonically by
increasing the slope. But this is only the case on a sufficiently
coarse grained scale, where $D(a)$ can in fact be qualitatively
matched to the results of two simple random walk models
\cite{RKD2}. On a fine scale, however, $D(a)$ shows a complicated
structure with different regions exhibiting different kinds of
self-similarity. A numerical estimation shows that the curve has a
fractal dimension which is very close to, but greater than one.

This highly irregular behavior of $D(a)$ is caused by correlations of
increasingly higher order in the microscopic dynamics of the map. For
instance, in the initial region $2\le a\le 3$, which has been
magnified in Fig.\ \ref{fig3}, the fine structure can be physically
explained by relating local extrema on the curve to characteristics of
the microscopic scattering process in one box as it changes with the
parameter $a$ \cite{RKD,RKDiss}: if stronger {\em backscattering} sets
in by making $a$ larger, the curve exhibits a local maximum, if
stronger {\em forward scattering} occurrs, it goes through a local
minimum.

More generally, the fractal character of $D(a)$ can be understood by
analyzing the Green-Kubo Eq.\ (\ref{eq:gk}) \cite{RKDiss,RKDG}. Two
basic components in the formula are responsible for the fractal
character of the curve: On the one hand, the diffusion coefficient is
given in terms of sums of jumps, which, according to Eq.\ (\ref{DTa}),
are related to fractal generalized Takagi functions as shown in Fig.\
\ref{fig2} (a) and (b). This goes together with the jump velocity
$j_a(x)$ having a discontinuity which varies with the parameter $a$
and which reveals in a sense the fractal character of the generalized
Takagi functions. On the other hand, a second source of irregularity
are the stepwise discontinuities in the density of the invariant
measure $\varrho_a(x)$ as shown in Fig.\ \ref{fig2} (c) and (d). The
irregular behavior of the diffusion coefficient results from a
combination of these effects, which are connected in the Green-Kubo
formula via integrating the respective generalized Takagi functions
over the respective invariant density. Thus, actually this behavior
finds its origin in the non-robustness of the topology of the
trajectories under parametric perturbations.

\subsubsection{A time-reversible area-preserving multibaker with fractal
diffusion coefficients}
\begin{figure}
\epsfxsize=9cm
\centerline{\epsfbox{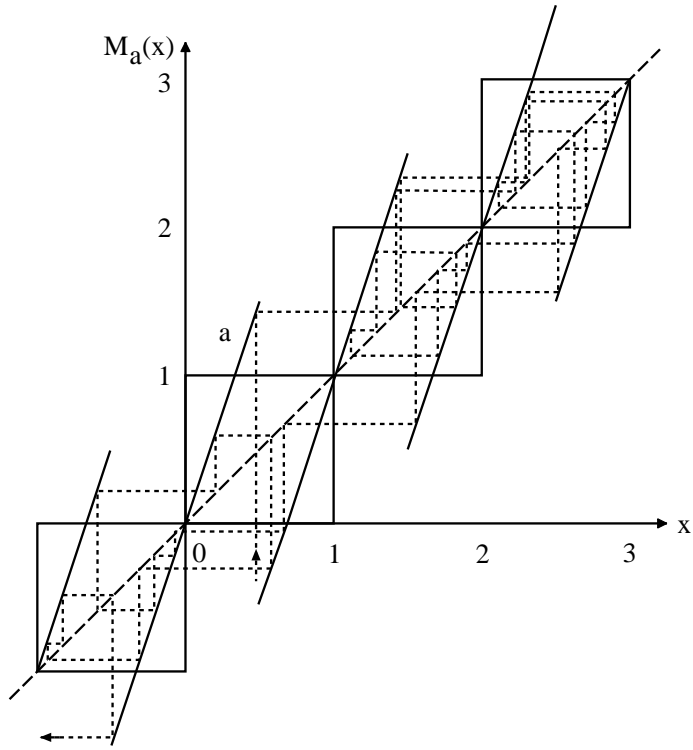}}
 
\vspace*{0.5cm}
\caption{\label{fig1} A simple model for deterministic diffusion. The
slope $a$, here $a=3$, serves as a control parameter in the
periodically continued piecewise linear map.}
\end{figure}

\begin{figure}
\epsfxsize=10cm
\centerline{\rotate[r]{\epsfbox{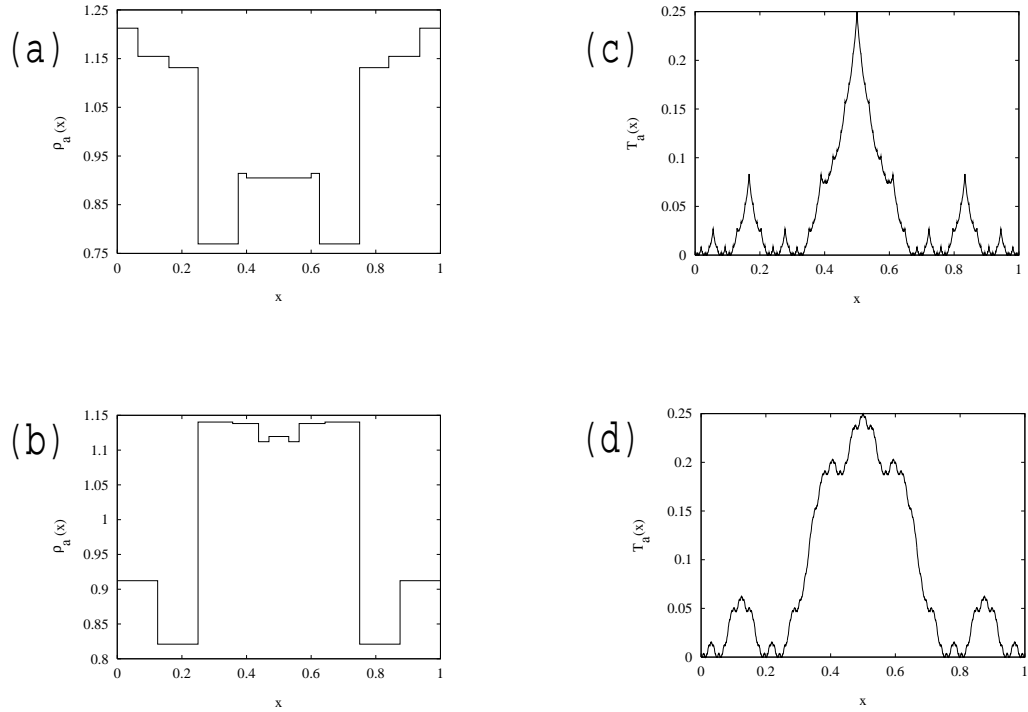}}}
\caption{\label{fig2} (a), (b) Invariant probability densities
$\varrho_a(x)$ on the unit interval for the map of Fig.\ \ref{fig1}
modulo 1. The slope is $a\simeq 2.5004$ for (a) and
$a\simeq 3.49997$ for (b). (c), (d) Generalized Takagi functions
$T_a(x)$ for the same map at $a=3$ in (c) and at $a=4$ in (d).}
\end{figure}

\begin{figure}
\epsfysize=11cm
\centerline{\rotate[r]{\epsfbox{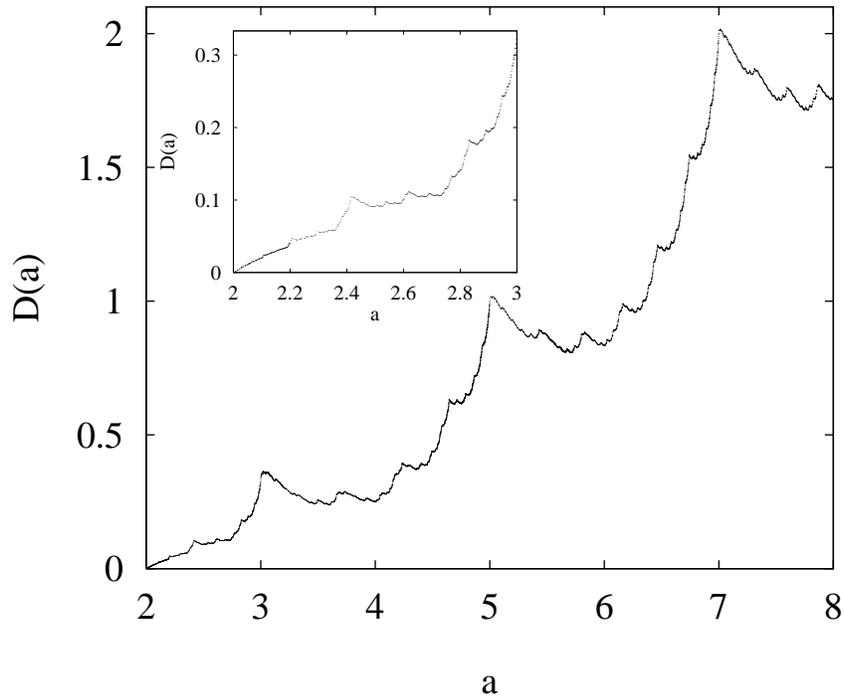}}}
  
\vspace*{0.4cm}
\caption{\label{fig3}
Parameter-dependent diffusion coefficient $D(a)$ for the map of Fig.\
\ref{fig1} and blowup of the initial region. The main graph consists of
7,908 single data points, the magnification of 979. In both cases
errorbars are too small to be visible.}
\end{figure}

\begin{figure}
\epsfysize=17cm
\centerline{\rotate[r]{\epsfbox{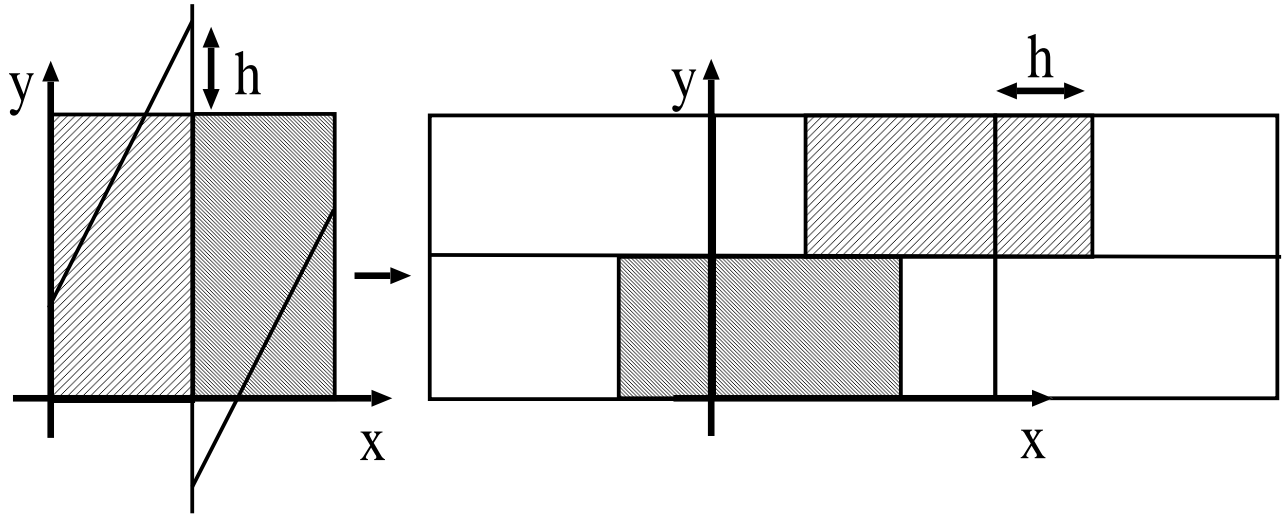}}}
  
\vspace*{1cm}
\caption{\label{fig5}
Dynamics of one cell of an area-preserving time-reversible
multibaker with a non-trivial parameter-dependence $h$. Projection 
of the dynamics onto the horizontal axis reduces the system to the
symmetric one-dimensional piecewise linear map shown in the figure to
the left which for $h=0$ is the Bernoulli shift.}
\end{figure}

The same phenomenon of a fractal diffusion coefficient appears in a
parameter-dependent generalization of the diffusive-reactive
multibaker model introduced above \cite{RKT}. This two-dimensional
area-preserving map is sketched in Fig.\ \ref{fig5}. Here, the two
rectangles of the left and of the right half of the square are
``sliding'' along the upper and the lower horizontal channel of the
periodically continued map governed by a parameter $h$, as shown in
the figure. It should be noted that for $h=0.5$ and shifting the
coordinate system by $\Delta x=0.5$ the model reduces to the simple
dyadic multibaker of Eq.\ (\ref{multibaker}). The dynamics of the
probability density $\tilde{f}_t(x,y,l)$ of the full multibaker
$\bPhi_h(x,y,l)$ is determined by the Frobenius-Perron equation
$\tilde{f}_{t+1}(x,y,l)=\tilde{f}_{t}\lbrack
\bPhi_h^{-1}(x,y,l)\rbrack$, where $\bPhi_h^{-1}(x,y,l)$ is the
inverse map. A projection of this two-dimensional Frobenius-Perron
equation onto the unstable $x$-direction by integrating over the
stable $y$-direction via $\varrho_t(x,l)\equiv \int dy
\:\tilde{f}_t(x,y,l)$ \cite{Dorfman,PG1} shows that the dynamics of
the probability density $\varrho_t(x,l)$ is determined by the
Frobenius-Perron equation of the simple one-dimensional map included
in Fig.\ \ref{fig5}, which is a kind of Bernoulli map shifted
symmetrically by a height $h$. This one-dimensional map governs the
dynamics of the multibaker map projected on the $x$-axis.  By
extending the system periodically, we recover a chain of
one-dimensional maps of the type of the one shown in Fig.\ \ref{fig1}.
Concerning time-reversibility, we follow the definition that there must
exist an involution $\bf G$ in phase space, ${\bf G}\circ{\bf G}=1$,
which reverses the direction of time via ${\bf G}\circ\bPhi\circ{\bf
G}=\bPhi^{-1}$ \cite{RK2}. For the special case of $h$ taking
multiples of $1/2$ involutions $\bf G$ can be found which are related
to a simple mirroring in phase space
\cite{invol}. For general $h$, it can be shown that the system has
strong time-reversible properties, although the existence of an
involution $\bf G$ remains an open question \cite{RKT,RK2}.

To compute the parameter-dependent diffusion coefficient of this
multibaker we use that the projected dynamics is governed by a
one-dimensional map, and thus we apply the same methods as outlined
above. The result is shown in Fig.\ \ref{fig6}. The diffusion
coefficient is again a non-trivial function of the parameter $h$ and
shares many characteristics of the curve presented in Fig.\
\ref{fig3}, for example, a certain random walk-like behavior on a coarse
grained scale \cite{RKD2}. But it also exhibits some new features,
especially that the diffusion coefficient is constant in intervals
$0.5+m\le h\le 1+m\: , \: m\in \hbox{\mb N}_0$. This is due to the
fact that the transition matrices corresponding to respective Markov
partitions, and thus the respective symbolic dynamics of the map, do
not change in this parameter interval. It is worth mentioning that in
contrast to the specific model of Fig.\ \ref{fig1} the invariant
probability density of the projected one-dimensional map here is
always uniform for all parameter values of $h$. Therefore, the only
contributions to the fractality of $D(h)$ come from the inter-cell
dynamics as described by the Takagi functions $T_a(x)$.

Along the same lines as above, we can also consider parametric
variations of a bias in one-dimensional maps and in multibaker models
\cite{RKT,JG,RK1,TVB}. In these systems, the deterministic dynamics
appears in form of currents which are fractal functions of the bias,
in certain parameter regions the mean current can run opposite to the
bias, and the diffusion coefficient can be zero with non-zero current
\cite{RKT,JG,RK1}. In this regard, it is interesting to point out that
drift currents which are irregularly fluctuating by varying the bias
have also been observed numerically in other deterministic models
\cite{Dett}. Moreover, we notice that certain biased maps can be
related to so-called ratchets \cite{RK1,PJ}.

\section{Reactive dynamics}
\subsection{Reactive modes of the dyadic multibaker}

In this section, we turn to the study of the chemiodynamic or reactive
modes of our simple dyadic model Eq.\ (\ref{multibaker}) of
diffusion-controlled reaction. Contrary to the total number of
particles, $N_A+N_B$, which is a constant of motion, the numbers of
particles of each chemical species are not conserved. Accordingly, we
should not expect that the reactive modes have a vanishing decay rate
as $k\to 0$. This is in contrast to the diffusive modes which are
related to the conserved total number of particles and for which the
decay rate\ (\ref{disp}) vanishes at $k=0$.

Here, we consider the subspace defined by the difference between the
particle concentrations in the multibaker,
\begin{equation}
g(x,y,l) \ \equiv \ f(x,y,l,A) \ - \ f(x,y,l,B) \ . \label{eq:pcm}
\end{equation}
Thus, we employ the fact that the dynamics of the concentration
difference can be decoupled from the mean density for this model, as
has been mentioned before, compare to Eq.\ (\ref{difference}).

With quasiperiodic boundary conditions, the difference of chemical
concentration evolves in time according to the reactive evolution
operator
\begin{equation}
\hat R_k \equiv \ \cases{
g_{t+1}(x,y,0) \ = \ \theta\Bigl(\frac{1}{2}-y\Bigr) \
g_t\bigl(\frac{x}{2},2y,1\bigr) \ + \ e^{-ikL} \
\theta\Bigl(y-\frac{1}{2}\Bigr) \ g_t\bigl(\frac{x+1}{2},2y-1,L-1\bigr)\ , \cr
g_{t+1}(x,y,1) \ = \ \theta\Bigl(\frac{1}{2}-y\Bigr) \
g_t\bigl(\frac{x}{2},2y,2\bigr) \ - \ \theta\Bigl(y-\frac{1}{2}\Bigr) \
g_t\bigl(\frac{x+1}{2},2y-1,0\bigr)\ , \cr
g_{t+1}(x,y,2) \ = \ \theta\Bigl(\frac{1}{2}-y\Bigr) \
g_t\bigl(\frac{x}{2},2y,3\bigr) \ + \ \theta\Bigl(y-\frac{1}{2}\Bigr) \
g_t\bigl(\frac{x+1}{2},2y-1,1\bigr)\ , \cr
\qquad\qquad\vdots \cr
g_{t+1}(x,y,L-2) \ = \ \theta\Bigl(\frac{1}{2}-y\Bigr) \
g_t\bigl(\frac{x}{2},2y,L-1\bigr) \ + \ \theta\Bigl(y-\frac{1}{2}\Bigr) \
g_t\bigl(\frac{x+1}{2},2y-1,L-3\bigr)\ , \cr
g_{t+1}(x,y,L-1) \ = \ - \ e^{ikL} \ \theta\Bigl(\frac{1}{2}-y\Bigr) \
g_t\bigl(\frac{x}{2},2y,0\bigr) \ + \ \theta\Bigl(y-\frac{1}{2}\Bigr) \
g_t\bigl(\frac{x+1}{2},2y-1,L-2\bigr)\ . \cr} \label{reactFP}
\end{equation}
Our goal is here to obtain the eigenvalues and eigenstates of this
reactive evolution operator
\begin{equation}
\hat R_k \ \bigl\lbrace \Psi(x,y,l)\bigr\rbrace_{l=0}^{L-1} \ = \ e^{s(k)} \ 
\bigl\lbrace \Psi(x,y,l)\bigr\rbrace_{l=0}^{L-1} \ ,
\end{equation}
with $\chi(k)=\exp\lbrack s(k)\rbrack$. We define the cumulative
functions
\begin{equation}
G_t(x,y,l) \ = \ \int_0^x \: dx' \: \int_0^y \: dy' \ g_t(x',y',l) \ ,
\end{equation}
which obey a set of equations which can be derived from Eq.\
(\ref{reactFP}). We suppose that the leading eigenstates are uniform
along the unstable direction $x$, which is justified by the fact that
the hyperbolic dynamics smoothens out any heterogeneities along the
unstable direction,
\begin{equation}
\Psi(x,y,l) \ = \ {\cal D}(y,l) \ ,
\end{equation}
where ${\cal D}(y,l)$ is a Schwartz distribution. We note that the
further eigenstates and root states do depend on $x$ and require a
more detailed analysis. The cumulative functions of the leading
eigenstates are thus
\begin{equation}
G_{\rm eigenstate}(x,y,l) \ \equiv \ x \ C(y,l) \qquad \hbox{with} \qquad
C(y,l) \ = \ \int_0^y \: dy' \ {\cal D}(y',l) \ .
\end{equation}
Replacing $g_t(x,y,l)$ by ${\cal D}(y,l)$ and $g_{t+1}(x,y,l)$ by
$\chi{\cal D}(y,l)$ in Eq. (\ref{reactFP}) and integrating over the
interval $\lbrack 0,y\rbrack$, we obtain the following iterative
equations for the new functions $C(y,l)$
\begin{equation}
\cases{
C(y,0) \ = \ \cases{ \frac{1}{2\chi} \ C(2y,1) \ , \qquad y<1/2 \ , \cr
\frac{1}{2\chi}\ \lbrack C(1,1)+ \exp(-ikL) C(2y-1,L-1)\rbrack \ ,\qquad y>1/2\
,\cr} \cr
C(y,1) \ = \ \cases{ \frac{1}{2\chi} \ C(2y,2) \ , \qquad y<1/2 \ , \cr
\frac{1}{2\chi}\ \lbrack C(1,2) - C(2y-1,0)\rbrack \ ,\qquad y>1/2\ ,\cr}
\cr
C(y,2) \ = \ \cases{ \frac{1}{2\chi} \ C(2y,3) \ , \qquad y<1/2 \ , \cr
\frac{1}{2\chi}\ \lbrack C(1,3) + C(2y-1,1)\rbrack \ ,\qquad y>1/2\ ,\cr}
\cr
\qquad\qquad\vdots\cr
C(y,L-2) \ = \ \cases{ \frac{1}{2\chi} \ C(2y,L-1) \ , \qquad y<1/2 \ , \cr
\frac{1}{2\chi}\ \lbrack C(1,L-1) + C(2y-1,L-3)\rbrack \ ,\qquad y>1/2\ ,\cr}
\cr
C(y,L-1) \ = \ \cases{ -\frac{\exp(ikL)}{2\chi} \ C(2y,0) \ , \qquad y<1/2 \ ,
\cr \frac{1}{2\chi}\ \lbrack - \exp(ikL) C(1,0) + C(2y-1,L-2)\rbrack \ ,\qquad
y>1/2\ .\cr} \cr} \label{iteration}
\end{equation}
The eigenvalue can be obtained by setting $y=1$ in
Eq. (\ref{iteration}), which leads to the eigenvalue equation
\begin{equation}
\pmatrix{-2\chi & 1 & 0 & 0 & \cdots & 0 & 0 & \exp(-ikL) \cr
-1 & -2\chi & 1 & 0 & \cdots & 0 & 0 & 0 \cr
0 & 1 & -2\chi & 1 &  \cdots & 0 & 0 & 0 \cr
0 & 0 & 1 & -2\chi &  \cdots & 0 & 0 & 0 \cr
\vdots & \vdots & \vdots & \vdots & \ddots & \vdots & \vdots & \vdots \cr
0 & 0 & 0 & 0 & \cdots & 1 & -2\chi & 1 \cr
-\exp(ikL) & 0 & 0 & 0 & \cdots & 0 & 1 & -2\chi\cr}\ 
\pmatrix{C(1,0) \cr C(1,1) \cr C(1,2) \cr C(1,3) \cr \vdots \cr C(1,L-2) \cr
C(1,L-1) \cr} \ = \ 0 \ . \end{equation} 
The characteristic determinant has been calculated for several values
of the distance $L$ between the reactive sites,
\begin{eqnarray}
L=3 &:& \qquad 4\chi^3 \ + \ \chi \ + \ \cos(3k) \ = \ 0 \ , \\
L=4 &:& \qquad 8\chi^4 \ - \ 2 \ + \ 2\cos(4k) \ = \ 0 \ , \\
L=5 &:& \qquad 16\chi^5\ -\ 4\chi^3\ -\ 3\chi \ + \ \cos(5k) \ = \ 0\ , \\ 
L=6 &:& \qquad 32\chi^6\ -\ 16\chi^4\ -\ 6\chi^2\ +\ 1 \ + \ \cos(6k) \ = \
0\ , \\ 
& & \qquad \qquad \qquad \vdots \nonumber
\end{eqnarray}
The corresponding dispersion relations of the reactive modes are depicted in
Fig.\ \ref{figDispersion} together with those of the diffusive modes.  Fig.\
\ref{figDispersion} shows that the slowest decay rate which gives the reaction
rate appears at $k=0$ for $L$ odd and at $k=\pm\pi/L$ for $L$ even.
The cumulative functions $\lbrace C(y,l)\rbrace_{l=0}^3$ of the
eigenstate corresponding to the reaction rate at $k=0$ are depicted in
Fig.\ \ref{figEigen} for the model with $L=3$ by solving Eq.\
(\ref{iteration}) iteratively. Near its maximum values, the dispersion
relation behaves quadratically like
\begin{eqnarray} 
L \ {\rm odd}&:& \qquad s^{({\rm r})}(k,L) \ =
\: - \: \tilde\kappa(L) \: - \: D^{({\rm r})}(L) \: k^2 \: + \: {\cal O}(k^4) 
\qquad  {\rm at} \quad k=0 \ , \\ 
L \ {\rm even}&:& \qquad s^{({\rm r})}(k,L) \ = \: - \:
\tilde\kappa(L) \: - \: D^{({\rm r})}(L) \ \bigl(k\mp\pi/L\bigr)^2 \: + \: 
{\cal O}\Bigl\lbrack \bigl(k\mp\pi/L\bigr)^4 \Bigr\rbrack \qquad 
{\rm at} \quad k=\pm
\pi/L \ .  \end{eqnarray} 
An analytical calculation of the reaction rate $\tilde\kappa(L)$ and a
numerical calculation of the reactive diffusion coefficient $D^{({\rm
r})}(L)$ versus $L$ reveal that
\begin{eqnarray}
\tilde\kappa(L) \ &=& \ - \ \ln \ \cos \frac{\pi}{L} \ = \ \frac{\pi^2}{2L^2} \
+ \ {\cal O}(L^{-4}) \ , \\
D^{({\rm r})}(L) \ &\sim& \ \frac{1}{L} \ .
\end{eqnarray}
The reaction rate thus behaves as expected for diffusion-controlled
reaction in one dimension, compare to Eq.\ (\ref{DiffControlReact}).
These results, combined with the results for the diffusive modes, show
that, on macroscopic scales, the coarse-grained density and the
concentration difference
\begin{eqnarray}
\rho(l) \ &=& \ \int_0^1 \: dx \: \int_0^1 \: dy \ \tilde{f}(x,y,l) \ , \\
\sigma(l) \ &=& \ \int_0^1 \: dx \: \int_0^1 \ dy \ g(x,y,l)\ = \ G(1,1,l) \ ,
\end{eqnarray}
behave like
\begin{eqnarray}
\hbox{diffusive \ mode} &:& \qquad \frac{\partial\rho}{\partial t} \
\simeq \ D \ 
\frac{\partial^2\rho}{\partial l^2} \ , \\
\hbox{reactive \ mode} &:& \qquad \frac{\partial\sigma}{\partial t} \ \simeq \
D^{(\rm r)} \ \frac{\partial^2\sigma}{\partial l^2} \ - \ \tilde\kappa
\ \sigma
\qquad (L\ {\rm odd}) \ . \label{reactivemode}
\end{eqnarray}
Corrections with higher-order spatial derivatives could also be taken
into account in the dynamics of the reactive mode. For a model with
$L$ odd, this behavior corresponds to a macroscopic diffusion-reaction
system with
\begin{eqnarray}
\frac{\partial\rho_A}{\partial t} \ &\simeq& \ \frac{D+D^{({\rm r})}}{2} \
\frac{\partial^2\rho_A}{\partial l^2} \ + \ \frac{D-D^{({\rm r})}}{2} \
\frac{\partial^2\rho_B}{\partial l^2} \ - \ \frac{\tilde\kappa}{2} \
(\rho_A-\rho_B) \ , \\
\frac{\partial\rho_B}{\partial t} \ &\simeq& \ \frac{D-D^{({\rm r})}}{2} \
\frac{\partial^2\rho_A}{\partial l^2} \ + \ \frac{D+D^{({\rm r})}}{2} \
\frac{\partial^2\rho_B}{\partial l^2} \ + \ \frac{\tilde\kappa}{2} \
(\rho_A-\rho_B) \ , 
\end{eqnarray}
where $\rho_A=\rho+\sigma/2$ and $\rho_B=\rho-\sigma/2$.  According to
these macroscopic equations, the diffusion coefficient of each species
is $D_A=D_B=(D+D^{({\rm r})})/2$, the cross-diffusion coefficient is
$D_{AB}=D_{BA}=(D-D^{({\rm r})})/2$, while the reaction rate of Eq.\
(\ref{reaction}) is given by the logarithm of the absolute value of
the leading eigenvalue of the reactive evolution operator as
\begin{equation}
\kappa \ = \ \frac{\tilde\kappa}{2} \ = \ - \ \frac{1}{2} \ \ln \ \cos
\frac{\pi}{L} \ = \ - \ \frac{1}{2} \ \ln \ \vert \chi(k=0)\vert \
. \label{reactionrate}\end{equation} 
We remark that, according to the microscopic analysis, the macroscopic
equations of a diffusion-reaction system do not necessarily follow the
simple assumption often carried out that the cross-diffusion
coefficients vanish, $D_{AB}=D_{BA}=0$.  This particular case is only
recovered if $D=D^{({\rm r})}$, which is not fulfilled here.  The
origin of this difference holds in the fact that the diffusion
coefficient $D_A$ associated with the state $A$ of a particle is in
general different from the diffusion coefficient of the particle
itself which may be in two possible states $A$ or $B$.  In this
regard, the cross-diffusion appears of importance in reacting systems.

Besides, the models with $L$ even follow more complicated
diffusion-reaction equations where the reactive diffusion coefficient
$D^{({\rm r})}$ has a different status because it is associated with a
nonvanishing wavenumber $k=\pm\pi/L$. Nevertheless, the part of the
diffusion-reaction process which is responsible for the reactive
exponential decay is confirmed by the microscopic analysis.

\subsection{Reaction rates in the parameter-dependent multibaker}
We now discuss the parameter-dependent reactive multibaker by taking
the shift parameter $h$ into account, as it has already been done for
the purely diffusive case (see Fig.\ \ref{fig5}). Thus, in addition to
the integer periodicity $L$ of the reaction cells of the multibaker
the reaction rate $\kappa$ will also depend on $h$. One may then raise
the question how the reaction rate $\kappa(h,L)$ changes with respect
to varying $h$ for fixed $L$. Moreover, we will give some illustrative
features of the time-dependent dynamics of the reaction process for
typical $h$ parameters.

Analogously to the previous subsection, we start with the difference
of chemical concentrations $g(x,y,l)$ as defined in Eq.\
(\ref{eq:pcm}). We again use the property that parallel to the
$x$-axis the two-dimensional reactive multibaker can be projected onto
a one-dimensional map, as has been pointed out before (see Fig.\
\ref{fig5}). The time evolution of the projected reactive part
$\zeta_t(x,l)\equiv \int dy\: g(x,y,l)$ of the multibaker is then
determined by the reactive evolution equation of a respective
one-dimensional map,
\begin{equation}
\zeta_t(x,l)=\hat R^{(1)}\:(h,L)\:\zeta_{t-1}(x,l) \ .
\label{eq:r1d} \end{equation}
Here, $\hat R^{(1)}(h,L)$ represents the one-dimensional reactive
evolution operator, and $\zeta_t(x,l)\equiv
\varrho_t(x,l,A)-\varrho_t(x,l,B)$ is the difference between $A$
and $B$-particle densities in the corresponding one-dimensional map.
As has been done for the purely diffusive case, we again write this
equation as a matrix equation, where instead of $\hat R^{(1)}(h,L)$ a
topological transition matrix $\hbox{\helv T}(h,L)$ acts onto a
particle density vector ${\underline{\bf\zeta}}_t$. The matrix
$\hbox{\helv T}(h,L)$ is structured such that in case of reactive
scattering centers the elements in the corresponding columns of the
matrix have a negative sign, and thus a particle changes color by
leaving a reaction cell. Otherwise, the matrix is the same as
discussed for the diffusive case.

We first discuss some details of the time evolution of the reactive
modes. By integrating over $\zeta_t(x,l)$ or its respective vector
representation we obtain the difference between the total number of
$A$ and $B$-particles at discrete time $t$ which is $\xi_t\equiv
\sum_l \int dx \:\zeta_t(x,l)$.  From the corresponding phenomenological
time-continuous reaction equation Eq. (\ref{reactivemode}) one would
expect that for the reactive multibaker $\xi_t$ decays
exponentially after a suitable coarse graining according to
$\xi_t=\xi_0 \exp(-\tilde\kappa t)$. If this is the case, we can
define the reaction rate of the reactive multibaker in analogy to the
phenomenological equation as $\kappa(h,L)=\tilde\kappa/2$.

To compute $\kappa(h,L)$ according to this definition, we solve the
matrix formulation of Eq.\ (\ref{eq:r1d}) by iterating the transition
matrices $\hbox{\helv T}(h,L)$ numerically. As an initial particle
density we choose $\zeta_0(x,l)$ to be uniform in one reactive cell of
the multibaker, which corresponds to having only $A$ particles in this
cell with the number of $B$ particles being locally zero, and we make
the chain long enough such that the evolving density is not affected
by boundary conditions. Fig.\ \ref{fig7} (a) and (b) give two typical
examples of $\zeta_t(x,l)$ for certain parameter values of $h$ after
$t=40$ iterations. They show how the ``perturbation'' $\zeta_0(x,l)$,
which is a local initial deviation from the equilibrium state
$\zeta_t(x,l)=0\:(t\to\infty)$, spreads out along the $x$-axis by
exhibiting a rather complex fine structure with oscillations around
zero. Fig.\ \ref{fig7} (c) and (d) contain half-logarithmic plots of
$|\xi_t|$ with respect to the discrete time $t$. These plots reflect a
different dynamical behavior of $\xi_t$ for different magnitudes of
the reaction rate. For $\kappa(h,L)$ close to zero, see the upper two
curves in Fig.\ \ref{fig7} (c), $\xi_t$ decays apparently
non-exponentially for small times $t$. Only for larger times it
eventually reaches exponential decay. Thus, the system shows that it
is close to states of the $h$ parameter where it is non-reacting. For
the lowest curve in Fig.\ \ref{fig7} (c), which corresponds to an
intermediate reaction rate, $\xi_t$ provides initially strong periodic
fluctuations. They are partly due to the complex deterministic
dynamics of the reactive baker in one cell of the chain, as has
already been observed and explained for a one-dimensional purely
diffusive case \cite{RKDiss}. Apart from such strong periodic
oscillations on a fine scale, in Fig.\ \ref{fig7} (d) $\xi_t$ exhibits
an interesting crossover between a fast decay for smaller times and a
slower decay for larger times, where again it approaches exponential
behavior. This may reflect the fact that for larger reaction rates
$\kappa(h,L)\gg 0$ the reaction is controlled by the diffusive
dynamics. These features of $\xi_t$ should be compared to the
characteristics of the respective probability densities $\zeta_t(x,l)$
in the figure.

To obtain quantitative values for the reaction rate, Eq.\ (\ref{eq:r1d})
has been analyzed by solving the eigenvalue problem of the
corresponding transition matrix $\hbox{\helv T}(h,L)$, analogously to
what has been done in the previous subsection for the dyadic reactive
multibaker. In general, the spectra of $\hbox{\helv T}(h,L)$ are
extremely complicated \cite{otherpaper}. However, as has been argued
in the previous subsection for special cases, and supported by our
observation of long-time exponential decay of $\xi_t$ for the general
case, we expect that, in the limit of infinite time, the reaction in
the multibaker will always be governed by the slowest eigenmodes and
their respective eigenvalues. This motivates to define the
parameter-dependent reaction rate $\kappa(h,L)$ via the maximum of the
absolute value of the eigenvalues of $\hbox{\helv T}(h,L)$,
\begin{equation}
\kappa(h,L) \equiv  - \frac{1}{2}\ln|\chi_{\rm max}(h,L)|
\quad , \label{eq:kh} 
\end{equation}
analogously to Eq. (\ref{reactionrate}). Numerically, we find that for
large regions of the $h$ parameter a certain fundamental domain $L_F$
of the multibaker is sufficient to obtain the correct leading
eigenvalue $\chi_{\rm max}(h,L)$. This domain must always include
multiples of two reactive centers, and its length is defined by the
number $L_F\equiv 2L\:{\rm Int}(1+h)$ of cells of the multibaker. In
these regions, solutions for the eigenvalue problem of $\hbox{\helv
T}(h,L)$ defined on the domain $L_F$ lead to a maximum eigenvalue
$\chi_{\rm max}(h,L)$ as obtained by solving the corresponding
eigenvalue problem for longer and longer chain lengths $mL\to\infty\:
, \: m \in \hbox{\mb N}$. However, especially for small $L$ and large
$\kappa(h,L)$ this fundamental domain only provides an approximation
to the exact results which are then obtained by making the chain
length $mL$ large enough such that the error in $\chi_{\rm max}(h,L)$
with respect to $mL$ is sufficiently small. Fig.\ \ref{fig8} shows
some typical largest eigenmodes $\psi(x,l)$ on the fundamental domain
in cases where it gives the correct corresponding largest eigenvalue
$\kappa(h,L)$. For large reaction rates, to a certain respect the
largest eigenmodes behave like sine functions, see Fig.\ \ref{fig8}
(a), whereas for smaller reaction rates, the largest eigenmodes
approach two-periodic step-like functions as shown in Fig.\ \ref{fig8}
(b). Fig.\ \ref{fig8} (c) depicts the largest eigenmodes for a
parameter value of $h$ which is just at the borderline of a
non-reacting $h$ region, but where the system is nevertheless already
highly reactive. Here, the eigenmodes appear to be especially
complicated \cite{otherpaper}.

\begin{figure}
\epsfxsize=7.5cm
\centerline{\rotate[r]{\epsfbox{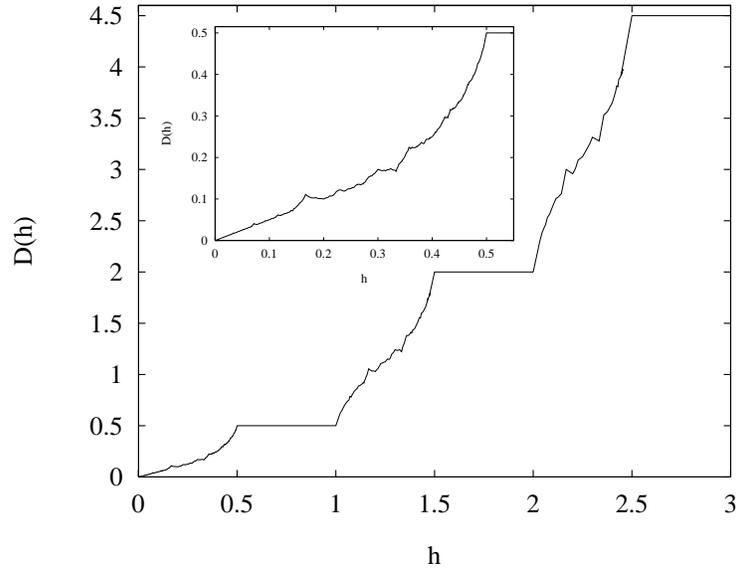}}}
  
\vspace*{0.5cm}
\caption{\label{fig6}
Parameter-dependent diffusion coefficient $D(h)$ for the multibaker of Fig.\
\ref{fig5} and blowup of the initial region. The main graph consists
of 638 data points, the magnification of 514. In both cases the single
points have been connected with lines, errorbars are too small to be
visible.}
\end{figure}

\begin{figure}
\epsfxsize=18cm
\centerline{\epsfbox{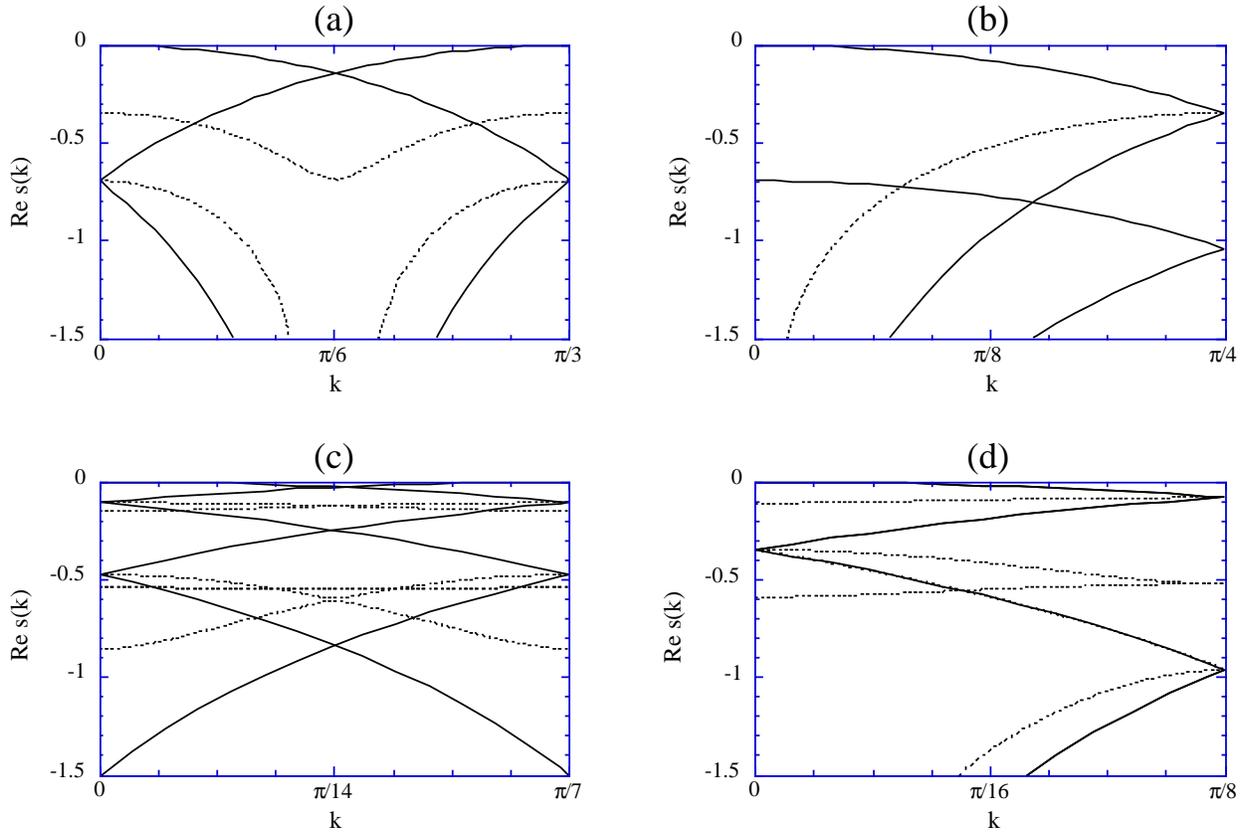}}
 
\vspace*{0.2cm}
\caption{\label{figDispersion} Dispersion relations of the diffusive
modes (solid lines) and of the reactive modes (dotted lines) for the
dyadic reactive multibaker with (a) $L=3$; (b) $L=4$; (c) $L=7$; (d)
$L=8$. {\em (note: poor quality output; the originals can be obtained
upon request)}}
\end{figure}

\begin{figure}
\epsfxsize=12cm
\centerline{\epsfbox{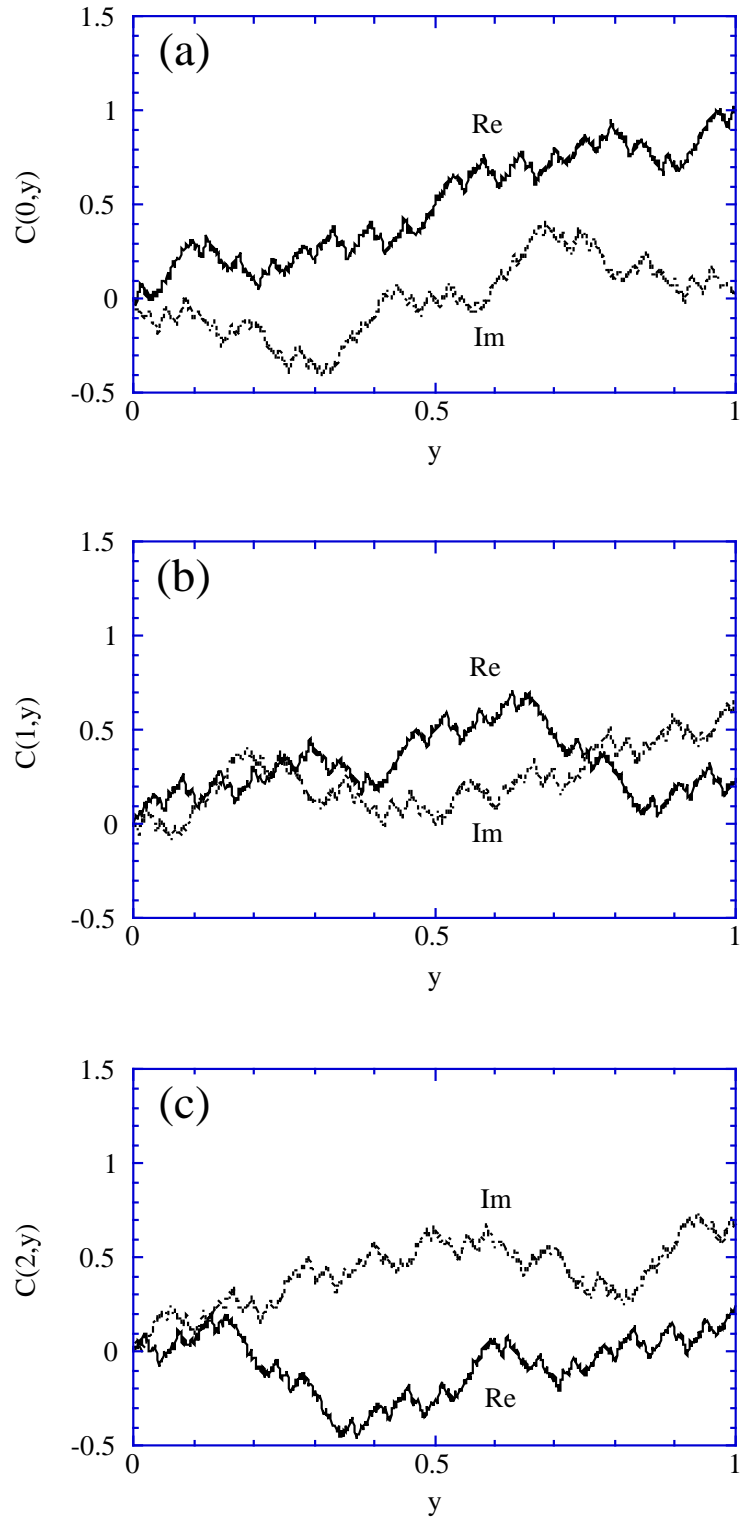}}
 
\vspace*{0.2cm}
\caption{\label{figEigen} Cumulative functions $\lbrace C(y,l)\rbrace$ with
$l=0,1,2$ of the reactive eigenstate at vanishing wavenumber $k=0$ in
the reactive multibaker model $L=3$. {\em (note: poor quality output;
the originals can be obtained upon request)}}
\end{figure}

\begin{figure}
\epsfysize=18cm
\centerline{\rotate[r]{\epsfbox{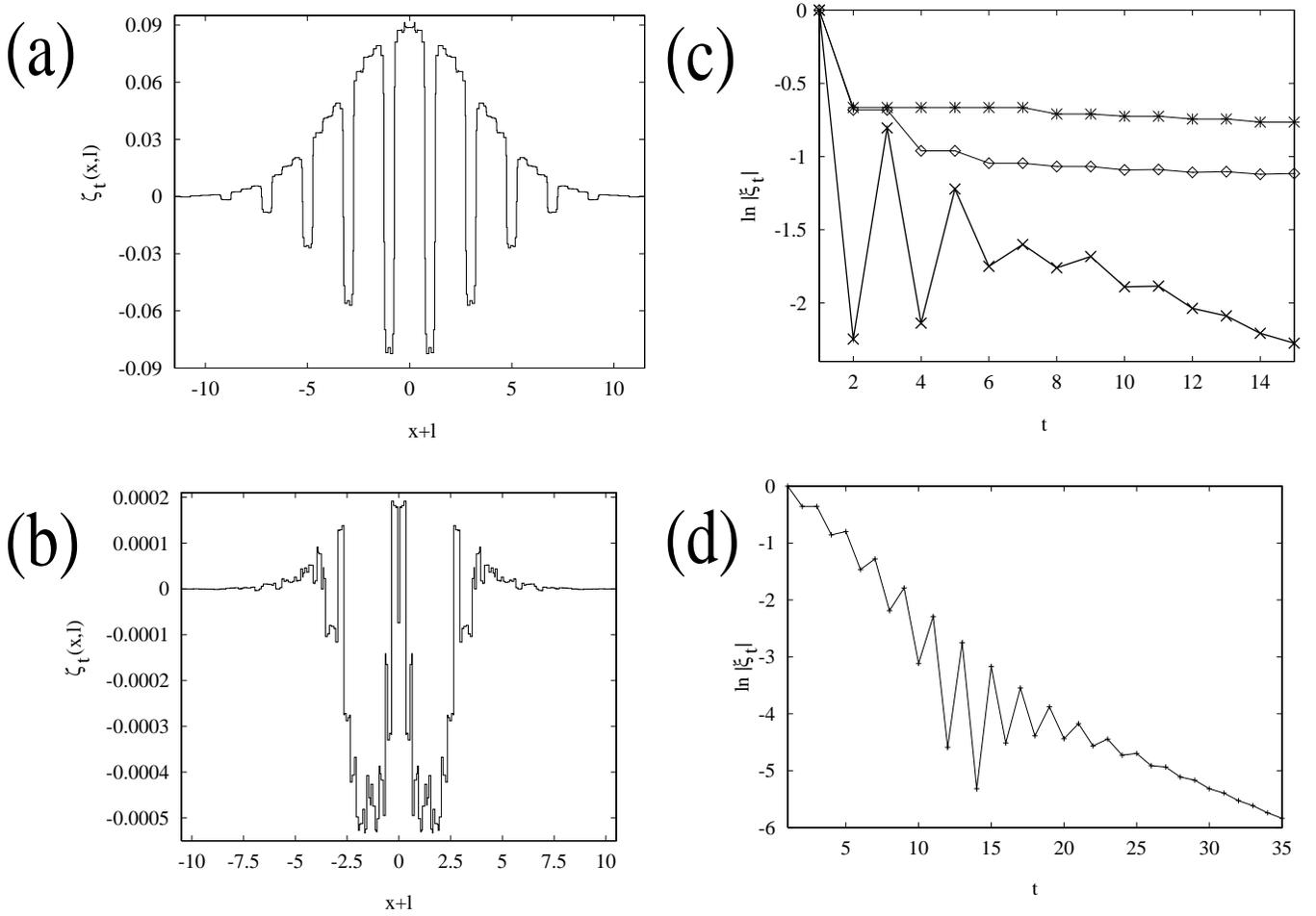}}}
 
\vspace*{0.5cm}
\caption{\label{fig7}(a), (b) Densities corresponding to the
difference in the number of $A$ and $B$-particles per total number of
particles after $t=40$ iterations of the map. In (a) ($L=2, h\simeq
0.2429$) the reaction rate is close to zero, whereas in (b) ($L=3,
h\simeq 0.1496$) it is locally maximal in $h$. (c), (d)
Half-logarithmic plots of the total difference $\xi_t$ in the number
of $A$ and $B$-particles of the system as it varies in time $t$. In
(c) the corresponding reaction rates $\kappa(h,L)$ are close to zero
for the upper two curves ($\kappa<0.006$), for the lower curve the
reaction rate has an intermediate value ($\kappa\simeq 0.02$), whereas
in (d) it is locally maximal in $h$ ($\kappa \simeq 0.05$). The
parameters for the upper curve in (c) correspond to (a), the curve in
the middle is at $L=3, h\simeq 0.247$, and the lowest one is at $L=3,
h\simeq 0.4472$. Case (d) corresponds to (b).}
\end{figure}

\begin{figure}
\epsfysize=20cm
\centerline{\epsfbox{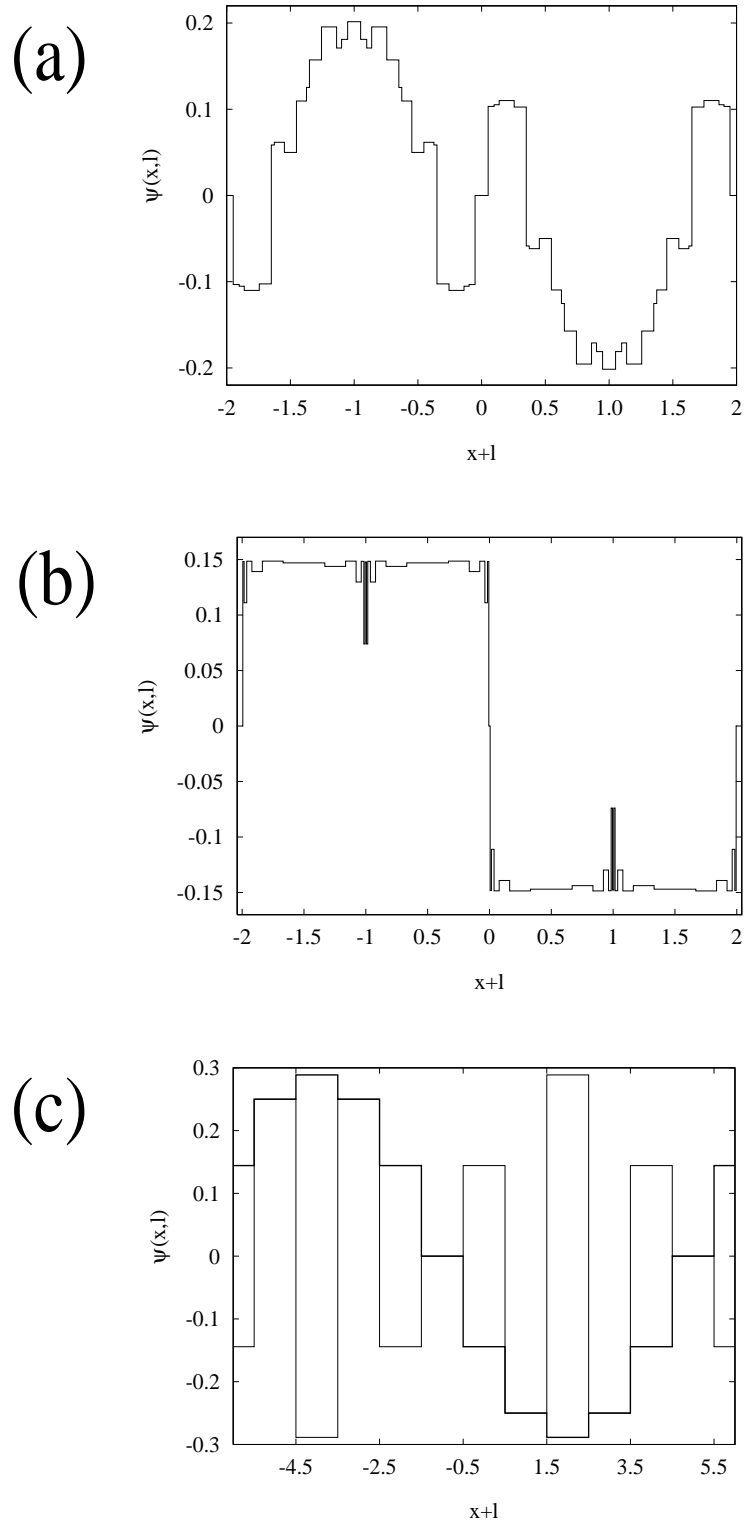}}
  
\vspace*{0.5cm}
\caption{\label{fig8} Examples of largest eigenmodes $\psi(x,l)$ for
the reactive multibaker corresponding to the largest eigenvalue
$\chi_{\rm max}(h,L)$ parallel to the $x$ axis in the fundamental
domain $L_F$ as described in the text. For (a) ($L=2, h\simeq 0.1496$)
the reaction rate $\kappa(h,L)$ is very large ($\kappa\simeq 0.12$),
for (b) ($L=2, h\simeq 0.4947$) it is very small ($\kappa \simeq
0.003$). In both cases, there exist only two real largest eigenmodes
where the second ones are shifted by a phase. In (c) ($L=6, h=1$) both
largest eigenmodes have been plotted (thick line for the one and thin
line for the other, respectively).}
\end{figure}

\begin{figure}
\epsfysize=20cm
\centerline{\epsfbox{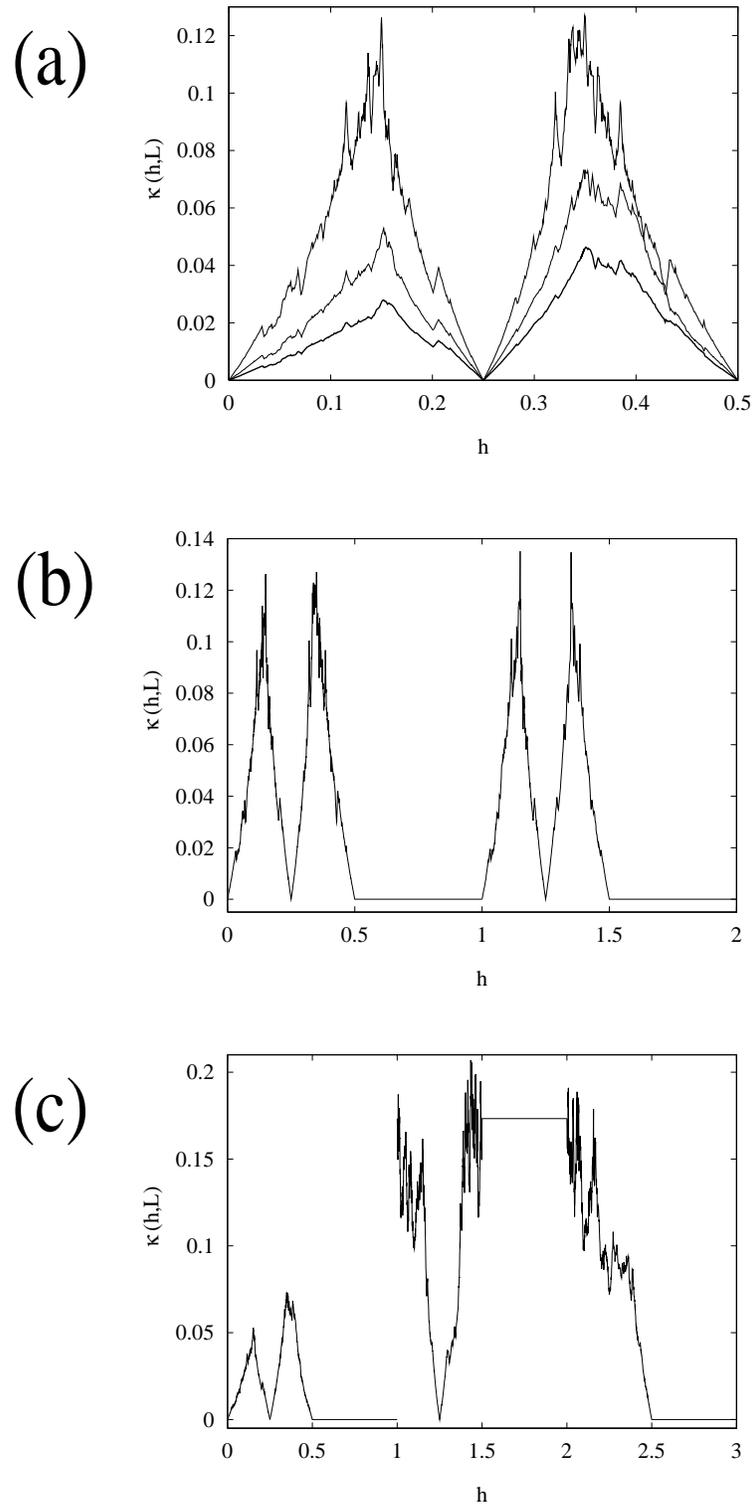}}
 
\vspace*{0.5cm}
\caption{\label{fig9}(a) Reaction rate $\kappa(h,L)$ at different
integer values of the reaction cell periodicity: $L=2$ (upper curve),
$L=3$ (middle), $L=4$ (lower curve). (b) Reaction rate for $0\le h\le
2$ at $L=2$. (c) Reaction rate for $0\le h\le 3$ at $L=3$.  In all
cases, error bars are too small to be visible.}
\end{figure}

In Fig.\ \ref{fig9} (a) the reaction rate $\kappa(h,L)$ as defined via
Eq.\ (\ref{eq:kh}) has been computed for a series of reaction center
periodicities $L$. For $h=0.25$ there is no reaction rate in the
system. In this case, the iteration method confirms that the
difference in the number of particles $\xi_t$ oscillates periodically
around zero instead of decaying exponentially. An analysis of the
eigenvalue spectra of the corresponding transition matrices reveals
that at this $h$ parameter the respective reactive multibakers are not
ergodic \cite{otherpaper}. For all other $h$ of the figure the
reaction rate is well-defined and shows a complicated structure. By
increasing $L$ the reaction rate $\kappa(h,L)$ decreases almost
everywhere, as one can expect intuitively, except in certain small
parameter regions of $h$. Fig.\ \ref{fig9} (b) gives the full result
for $\kappa(h,L)$ at $L=2$. This structure repeats itself with a
periodicity of $2m\le h\le 2+2m\: , \: m\in \hbox{\mb N}_0$. In
certain intervals of $h$ the four peaks depicted in the figure are
very similar, or even identical, however, there does not appear to be
a simple scaling law by which the full peaks can be mapped onto each
other. The plateau regions with zero reaction rate correspond to the
respective regions observed in Fig.\ \ref{fig6} for the diffusion
coefficient of the system. They share the same characteristics as
discussed above for the singular case of $h=0.25$, except that at
$h=1$ the system is ergodic, but not mixing. Topologically, these
regions are of the same origin as explained for the diffusive case. In
Fig.\ \ref{fig6} (c) the reaction rate has been computed for $L=3$. In
contrast to the two-periodic case the change from a non-reactive
region to a reactive region occurs for $L=3$ apparently
discontinuously in the reaction rate by varying $h$ around $1$. This
corresponds to the system at $h=1$ and $L=3$ being mixing, whereas for
the same $h$ and $L=2$ it was ergodic, but not mixing. We note that
for $L=4$ there are even two of such discontinuous transitions. The
detailed irregular structure of the curves, as well as the
phase-transition like behavior shown in Fig.\
\ref{fig9} (c), can be understood in more detail by analyzing the
eigenvalue spectra of the reactive evolution operator and how
they change under parameter variation, as will be discussed elsewhere
\cite{otherpaper}.

Apart from varying $h$, other parameter dependencies can be studied in
this reactive multibaker as well. For example, the distance between
the single reaction centers could be changed by allowing $L$ to be
continuous, the positions of the reaction centers could be shifted by
keeping $L$ fixed, and the size of the reaction centers could be
increased or decreased. In all these cases we expect non-trivial
parameter dependences to be typical which are similar to the one
depicted in Fig.\ \ref{fig9} \cite{otherpaper}.

\section{Discussion and Conclusions}
In this paper, we have analyzed simple deterministic models of
diffusion-controlled reaction. The models fulfill the chaotic
hypothesis mentioned in the Introduction, which allows us a much
sharper analysis of the phenomenological foundation of
diffusion-reaction processes than with the old stochastic assumption.
In this regard, for the simplest model we have been able to derive the
exact dispersion relations not only of the diffusive modes but also of
the reactive modes. The reactive modes indeed decay exponentially, as
it should be for nonconserved quantities.  The reaction rate behaves
like $\kappa \sim D/L^2$, as expected for diffusion-controlled
reactions in one dimension.  The reaction rate introduces a new time
scale $T_{\rm react}\sim \kappa^{-1}\sim L^2/D$, with respect to
purely diffusive systems.  This reactive time depends on the
concentration $\sim 1/L$ of catalysts and, therefore, it takes a fixed
value in a given diffusion-reaction system.  This time scale of
reaction is intermediate between the short time scale of the Lyapunov
instability $T_{\rm chaos}\sim \lambda_{\rm max}^{-1}$ and the long
time scale of diffusion $T_{\rm diff}\sim {\cal L}^2/D$.  An important
difference between diffusion and reaction is that the evolution
operator is positive in the diffusive subspace and is thus of
Frobenius-Perron type, although the evolution operator (\ref{reactFP})
has both signs in the reactive subspace.  By analyzing this
deterministic evolution operator, we have found that the spatial
dynamics of the reactive modes appears significantly different from
the standard assumption of macroscopic diffusion-reaction models due
to the particular importance of cross diffusion.  On the other hand,
the eigenstates associated with the reactive and diffusive modes are
expressed as singular Schwartz distributions, also in contrast with
the phenomenological models which have always suggested that the
eigenstates are regular functions. In a sense, we may say that this
singular character of the exact eigenmodes renders their relaxation
compatible with the deterministic dynamics of the particles, in full
respect to the mechanical Liouville theorem of volume preservation.
This result previously observed for diffusion is here shown to hold
also for reaction.

Moreover, the macroscopic transport coefficients such as the diffusion
coefficient and the reaction rate both turn out to exhibit a highly
irregular behavior as a function of a control parameter of the
system. In this way, we have pursued the early work by Elskens and
Kapral who studied the isomerization rate for the simplest values of
their model parameter \cite{Elskens}. The irregular behavior has its
origin in the topological instability of the trajectories in phase
space and also in the singular character as represented by Takagi
functions. It appears as a fundamental result because the topological
instability of phase-space dynamics is a common feature to many
systems including the Lorentz gas, the hard-sphere gas, and
nonhyperbolic systems which are all non-robust under parametric
perturbations. However, we may expect that the transport coefficients
would have an irregular behavior only if the systems fulfill certain
additional criteria, as being spatially periodic, being sufficiently
low-dimensional, and being such that particle-particle interactions
are not of main importance.  Physical systems of this kind could - to
a certain extent - already be realized experimentally in form of
so-called antidot lattices \cite{Weis91}. On the other hand, the
detailed fractal character of parameter-dependent transport
coefficients may disappear by increasing the number of degrees of
freedom, or by including strong stochastic perturbations
\cite{rp}. We have moreover observed that the reaction rate is not only
a highly irregular function of the parameter but also presents
discontinuities which are reminiscent of nonequilibrium phase
transitions.  The parametric sensitivity seems thus enhanced at the
level of the reactive properties as compared with the diffusive ones.

\vskip 0.2 cm

\noindent{\bf Acknowledgements.} The authors want to thank Prof.\ G.\
Nicolis for continuous support and encouragement. They also want to
thank Prof.\ T.\ T\'el for inviting them to the summer school/workshop
``Chaos and Irreversibility'' (E\"otv\"os University, Budapest,
August/September 1997) and for his hospitality. P.G.\ is financially
supported by the National Fund for Scientific Research (F.~N.~R.~S.\
Belgium). This research is partly financially supported by the
IUAP-PAI Program of the Belgian Federal Office SSTC and by the ``Human
Capital and Mobility Program'' of the European Commission. R.K.\
gratefully acknowledges a grant from the Deutsche
Forschungsgemeinschaft (DFG).

\newpage 

\end{document}